\documentclass[aps,prc,superscriptaddress,twocolumn,floatfix]{revtex4}
\usepackage{epsfig}
\usepackage[colorlinks=true,linktocpage=true,linkcolor=blue,citecolor=blue]{hyperref}
\usepackage[usenames,dvipsnames]{color}
\usepackage{amsmath, amssymb}
\usepackage{multirow}
\usepackage{longtable}
\usepackage{color}
\usepackage[normalem]{ulem}  

\renewcommand\sout{\bgroup \color{blue} \ULdepth=-.5ex \ULset}

\newcommand{\I}{{\rm I}}
\newcommand{\T}{{\rm T}}

\begin{document}

\title{Kinetic freeze out conditions in nuclear collisions with $2-158 $A GeV beam energy within a non boost-invariant blast wave model}

\author{Sudhir Pandurang Rode}
\affiliation{Discipline of Physics, School of Basic Sciences, Indian Institute of Technology Indore, Indore 453552 India}

\author{Partha Pratim Bhaduri}
\affiliation{Variable Energy Cyclotron Centre, HBNI, 1/AF Bidhan Nagar, Kolkata 700 064, India}

\author{Amaresh Jaiswal}
\affiliation{School of Physical Sciences, National Institute of Science Education and Research, HBNI, Jatni-752050, Odisha, India}

\author{Ankhi Roy}
\affiliation{Discipline of Physics, School of Basic Sciences, Indian Institute of Technology Indore, Indore 453552 India}

\date{\today}

\begin{abstract}

We study the kinetic freeze-out conditions of bulk hadrons in nuclear collisions. The transverse and longitudinal momentum spectra of the identified hadrons produced in central Au+Au and Pb+Pb collisions, in the beam energy range of $\rm E_{Lab}=2-158 $A GeV are analyzed for this purpose, within a generalized non boost-invariant blast wave model. The kinetic freeze-out temperature is found to vary in the range of $55-90$ MeV, whereas the average transverse velocity of collective expansion is found to be around $0.5c$ to $0.6c$. The mean longitudinal velocity of the fireball is seen to increase monotonically with increasing longitudinal boost. The results would be useful to understand the gross collision dynamics for the upcoming experiments at the FAIR and NICA accelerator facilities.
\end{abstract}

\maketitle


\section{Introduction}
\label{}

One of the major objectives of the relativistic heavy-ion collision research program is to explore the phase structure of the strongly interacting matter~\cite{Florkowski:2014yza, UWHeinz, BraunMunzinger:2008tz}. Regions of temperature and baryon density that can be accessed in a particular experiment, depending on the collision energy. Thus systems with very small net baryon densities but rather high temperature are formed at top Relativistic Heavy Ion Collider (RHIC) and Large Hadron Collider (LHC) energies. Data collected by the experiments at these two collider facilities~\cite{rhic1, rhic2, lhc1, lhc2, lhc3} have provided conclusive evidence for the formation of strongly coupled Quark Gluon Plasma (QGP)~\cite{sQGP}. Compared to this, nuclear phase diagram is much less explored in the region of high net baryon densities. Relativistic nuclear collisions at moderate energies such as those available at RHIC Beam Energy Scan (BES) program, at the upcoming FAIR accelerator facility at GSI Germany~\cite{CBM-physics} and NICA facility at JINR Dubna~\cite{nica}, are expected to create hot and dense nuclear matter in the regime of moderate temperature and large net baryon density. 

For optimum utilization of these facilities to explore the QCD phase structure, it is imperative to analyze the existing data sets in the similar energy range collected by the fixed target experiments at AGS and SPS accelerator facilities. In particular, it is important to know the thermodynamic conditions that are created in the bulk of the colliding system at various collision energies. Due to technological limitations related to beam intensities, the experiments are mostly limited to the measurement of soft probes that constitute the bulk of the excited matter produced in these collisions. Soft hadronic observables measure directly the final ``freeze-out'' stage of the collision when hadrons decouple from the bulk and free-stream to the detectors. They exhibit a strong collective behavior that can be analyzed within the ambit of relativistic hydrodynamics~\cite{hydro_review, hydro_SPS}. 

Apart from hydrodynamics, the observables pertaining to the collective behavior of the nuclear fireball can also be studied using the hydrodynamics inspired phenomenological model called the blast wave model~\cite{Florkowski:2004tn}. Due to its simplicity, blast wave models have been used for a long time to analyze momentum distribution of the produced hadrons and provide information about the properties of the matter at kinetic freeze-out. The main underlying assumption is that the particles in the system produced in the collisions are locally thermalized (till they are emitted from the medium) and the system expands collectively with a common radial velocity field undergoing an instantaneous common freeze-out. Such phenomenological models are particularly useful in nuclear collisions where a fireball is created at finite net baryon density because the hydrodynamic calculations in the corresponding regime suffer from the unavailability of the realistic equation of states from the lattice QCD.

The first version of blast wave model was formulated about four decades ago~\cite{Siemens}, to describe the hadron production in Ne+NaF reactions at a beam energy of 800A MeV. The model assumes the radial expansion of the fireball, with constant velocity. Collective isentropic expansion of the nuclear fireball with a scaling form for the radial velocity profile was also used to analyze the then available data on transverse momentum ($p_T$) spectra of hadrons from 14.5A GeV Si+Au collisions at BNL AGS and 200A GeV O+Au collisions at CERN SPS~\cite{Lee1, Lee2}. 

While the spherically expanding source may be expected to mimic the fireball created at low energies, at higher energies stronger longitudinal flow might lead to cylindrical geometry. For the latter case, an appropriate formalism was first developed in Ref.~\cite{Schnedermann}. Using a simple functional form for the phase space density at kinetic freeze out, the authors approximated the hydrodynamical results with the boost-invariant longitudinal flow. The model was successfully used to fit the $p_T$ spectra with only two parameters namely a kinetic freeze-out temperature $T_{kin}$ and a radial flow strength $\beta_T$. Though initially developed for central collisions, the model was later extended to non-central collisions with the introduction of additional parameters to account for anisotropies in the transverse flow profile~\cite{Pasi} and in the shape of the source in the co-ordinate space~\cite{RHIC130}. The model has also been applied to search for collectivity in small systems~\cite{prem}. Attempts have also been made to incorporate the viscous effects in the blast wave model~\cite{Teaney, Jaiswal:2017lph}. One common assumption for all these variants of the blast wave model is the underlying boost-invariant longitudinal dynamics. Although it is a reasonable assumption at RHIC and LHC energies, longitudinal boost-invariance does not hold well at AGS and SPS energies. Therefore in order to describe particle production at these energy domains, the assumption of boost-invariance must be relaxed. 

In Ref.~\cite{Dobler}, the authors proposed a non boost-invariant extension of the blast wave model of Ref.~\cite{Schnedermann}. The cylindrical symmetry is broken via the modification of the system boundaries which is suitable for low energy collisions. For a realistic parametrization of the freeze-out surface of the expanding fireball, the model has been found to provide a very good fit to the $p_T$ and rapidity spectra for a variety of hadrons produced in $11.6 $A GeV Au+Au collisions measured by E802, E877 and E891 Collaborations at AGS. The results indicated a relatively low kinetic freeze-out temperature of about $90$ MeV with an average transverse expansion velocity at mid-rapidity of about $0.5c$. In the present article, we follow the same prescription to analyze the transverse and longitudinal spectra of the bulk hadrons from Au+Au and Pb+Pb collisions in the energy range $\rm E_{Lab}=2-158$A GeV, as measured by different experimental Collaborations at AGS, SPS and RHIC facilities. The results of our current investigation would help to improve the understanding of the gross features of the collision dynamics in this energy domain which is certainly useful for the second generation experiments at FAIR and NICA accelerator facilities.

\begin{table*}[ht]\centering
\label{tabI}
\vglue4mm
\begin{tabular}{|c|c|c|c|c|c|c|c|} \hline
Facility & Experiment & E$_{\rm Lab}$~(A GeV) & $y_b$ & System & Centrality & Phase space & Hadron Species\\ \hline
AGS \rule{0pt}{0.5cm} & E895 & 2~\cite{Klay:2003zf} \cite{Klay:2001tf} & 1.39 & Au+Au &$0 - 5 \%$&$-0.05 < y_{\rm c.m.} < 0.05$ & $\pi^{+}$, $\pi^{-}$, $p$\\\hline
AGS \rule{0pt}{0.5cm} & E895 & 4 & 2.13 & Au+Au &$0 - 5 \%$&$-0.05 < y_{\rm c.m.} < 0.05$ & $\pi^{+}$, $\pi^{-}$, $p$\\\hline
AGS \rule{0pt}{0.5cm} & E895 & 6 & 2.54 & Au+Au &$0 - 5 \%$&$-0.05 < y_{\rm c.m.} < 0.05$ & $\pi^{+}$, $\pi^{-}$, $p$\\\hline
AGS \rule{0pt}{0.5cm} & E895 & 8 & 2.83 & Au+Au & $0 - 5 \%$&$-0.05 < y_{\rm c.m.} < 0.05$ & $\pi^{+}$, $\pi^{-}$, $p$\\\hline
SPS \rule{0pt}{0.5cm} & NA49 & 20~\cite{NA4920} & 3.75 & Pb+Pb &  $0 - 7 \%$&$0.0 < y_{\rm c.m.} < 0.2$ ($\pi^{-}$) & $\pi^{-}$, $\rm K^{\pm}$, $p$\\
 \rule{0pt}{0.5cm} &  &  &  &  & &$-0.1 < y_{\rm c.m.} < 0.1$ ($\rm K^{\pm}$) & \\
 \rule{0pt}{0.5cm} &  &  &  &  & &$-0.38 < y_{\rm c.m.} < 0.32$ ($p$) & \\
\hline
RHIC BES \rule{0pt}{0.5cm} & STAR & 30.67 ~\cite{Adamczyk:2017iwn} & 4.18 & Au+Au & $0 - 5 \%$&$-0.1 < y_{\rm c.m.} < 0.1$ & $\pi^{\pm}$, $\rm K^{\pm}$,$p$, $\bar{p}$\\\hline
SPS \rule{0pt}{0.5cm} & NA49 & 30 & 4.16 & Pb+Pb & $0 - 7 \%$&$0.0 < y_{\rm c.m.} < 0.2$ ($\pi^{-}$) & $\pi^{-}$, $\rm K^{\pm}$, $p$\\
 \rule{0pt}{0.5cm} &  &  &  &  & &$-0.1 < y_{\rm c.m.} < 0.1$ ($\rm K^{\pm}$) & \\
 \rule{0pt}{0.5cm} &  &  &  &  & &$-0.48 < y_{\rm c.m.} < 0.22$ ($p$) & \\
\hline
SPS \rule{0pt}{0.5cm} & NA49 & 40~\cite{NA4940} & 4.45 & Pb+Pb & $0 - 7 \%$&$0.0 < y_{\rm c.m.} < 0.2$ ($\pi^{-}$)& $\pi^{-}$, $\rm K^{\pm}$, $p$\\
 \rule{0pt}{0.5cm} &  &  &  &  & &$-0.1 < y_{\rm c.m.} < 0.1$ ($\rm K^{\pm}$) & \\
 \rule{0pt}{0.5cm} &  &  &  &  & &$-0.32 < y_{\rm c.m.} < 0.08$ ($p$) & \\\hline
RHIC BES \rule{0pt}{0.5cm} & STAR & 69.56 & 5.00 & Au+Au & $0 - 5 \%$&$-0.1 < y_{\rm c.m.} < 0.1$ & $\pi^{\pm}$, $\rm K^{\pm}$,$p$, $\bar{p}$\\\hline
SPS \rule{0pt}{0.5cm} & NA49 & 80 & 5.12 & Pb+Pb & $0 - 7 \%$&$0.0 < y_{\rm c.m.} < 0.2$ ($\pi^{-}$)& $\pi^{-}$, $\rm K^{\pm}$, $p$\\
 \rule{0pt}{0.5cm} &  &  &  &  & &$-0.1 < y_{\rm c.m.} < 0.1$ ($\rm K^{\pm}$) & \\
 \rule{0pt}{0.5cm} &  &  &  &  & &$-0.36 < y_{\rm c.m.} < 0.04$ ($p$) & \\\hline
SPS \rule{0pt}{0.5cm} & NA49~\cite{NA49158} & 160 & 5.82 & Pb+Pb & $0 - 7 \%$&$0.0 < y_{\rm c.m.} < 0.2$ ($\pi^{-}$) & $\pi^{-}$, $\rm K^{\pm}$, $p$\\
 \rule{0pt}{0.5cm} &  &  &  &  & &$-0.1 < y_{\rm c.m.} < 0.1$ ($\rm K^{\pm}$) & \\
 \rule{0pt}{0.5cm} &  &  &  &  & &$-0.51 < y_{\rm c.m.} < -0.11$ ($p$) & \\\hline
\end{tabular}
\caption{Details of the data sets from different experiments at different accelerator facilities along with energy ($\rm E_{Lab}$), beam rapidity ($y_b$), System, Centrality, Phase space and Hadron species, used for this blast wave analysis.}
\end{table*}


\section{A brief description of the model}

\begin{table*}[ht]\centering
\label{tabII}
\vglue4mm
\begin{tabular}{|c|c|c|c|c|} \hline
 $\rm E_{Lab}$~(A GeV) & $\eta_{max}$   & $\langle \beta_{T} \rangle$ & $T_{kin}$ (MeV) & $\chi^{2}$/NDF\\ \hline
2 \rule{0pt}{0.5cm} & $0.995 \pm 0.001$ &$0.4838 \pm 0.0034$ & $61.72 \pm 1.36$ & 7.2\\
 \hline
4 \rule{0pt}{0.5cm} & $1.285 \pm 0.002$  & 0.5400 $\pm$  0.0025 & 55.86  $\pm$  1.38 & 7.8\\
\hline
6 \rule{0pt}{0.5cm} & $1.573 \pm 0.002$ & 0.5584 $\pm$  0.0062 & 58.14  $\pm$  3.17 & 9.4\\
 \hline
8 \rule{0pt}{0.5cm} & $1.645 \pm 0.003$ & 0.5655 $\pm$  0.0031 & 60.63  $\pm$  1.75 & 8.7\\
\hline
20 \rule{0pt}{0.5cm} & $1.882 \pm 0.005$ &$0.5177 \pm 0.0011$ & $79.77 \pm 0.05$ & 6.5\\\hline
30.67 \rule{0pt}{0.5cm} & $2.078 \pm 0.004$ & 0.5448 $\pm$  0.0002 & 71.25  $\pm$  0.02 & 8.5\\\hline
30 \rule{0pt}{0.5cm} & $2.084 \pm 0.004$ &$0.5368 \pm 0.0011$ & $80.28 \pm 0.05$ & 6.7\\\hline
40 \rule{0pt}{0.5cm} & $2.094 \pm 0.004$ &$0.5356 \pm 0.0009$ & $81.92 \pm 0.04$ & 5.5\\\hline
69.56 \rule{0pt}{0.5cm} & $2.306 \pm 0.005$ & 0.5330 $\pm$  0.0001 & 78.97  $\pm$  0.01 & 6.7\\\hline
80 \rule{0pt}{0.5cm} & $2.391 \pm 0.005$ &$0.5347 \pm 0.0012$ & $82.68 \pm 0.05$ & 3.8\\\hline
158 \rule{0pt}{0.5cm} & $2.621 \pm 0.006$ &$0.538 \pm 0.0013$ & $84.11 \pm 0.06$ & 4.4\\\hline
\end{tabular}
\caption{Summary of the fit results at different energies from AGS, SPS and RHIC beam energy scan (BES). For uniformity, at RHIC the relevant centre of mass (CMS) energies are converted to the corresponding beam energies in the laboratory frame.}
\end{table*}

Details of the non boost-invariant blast wave model that we have employed in our calculations can be found in~\cite{Dobler}. Here we briefly outline the main features for completeness. In blast wave model, the single particle momentum spectrum of the hadrons emitted from the fireball at freeze-out is usually described by the Cooper-Frye~\cite{Cooper:1974mv} prescription of particle production. Within this formalism, the single particle spectrum is defined as the integral of the phase-space distribution function $f(x,p)$ over the freeze-out hyper-surface $\Sigma_{\mu}^{f}(x)$. The triple differential invariant spectra can be written as:
\begin{equation}
\label{CF}
E\frac{d^3N}{d^3p} = {\frac{g}{2 \pi^{3}}}\int d^{3}\Sigma_{\mu}^{f}(x)p^{\mu}f(x,p)
\end{equation}
where $g$ denotes the degeneracy factor. In thermal models, $f(x,p)$ is considered to be the equilibrium distribution function. In the temperature range of the heavy-ion collisions, the quantum statistics can be ignored and one usually works with the Boltzmann approximation. The freeze-out hypersurface $\Sigma^f_\mu(x)$ is determined from freeze-out criteria for thermal decoupling. 

For an expanding fireball in local thermal equilibrium, the boosted thermal distribution is given by:
\begin{equation}
\label{dist}
f(x,p) = \exp\left( -{\frac{p.u(x) - \mu(x)}{T(x)}} \right)
\end{equation}
where $T(x)$ and $\mu(x)$ are space-time dependent local temperature and chemical potential at kinetic freeze-out and $u^{\mu}(x)=\gamma (1,\beta_{T}(x)e_{r},\beta_{L}(x))$ is the local fluid velocity. Focusing on central collisions, with a realistic parametrisation of the freeze-out hyper-surface and local fluid velocity, the thermal single particle spectrum in terms of transverse mass $m_T (\equiv\sqrt{p_T^2+m^2})$ and rapidity $y$ are given by
\begin{eqnarray}
\label{therm}
   \frac{dN}{m_{T} dm_{T}  dy}
   &=&
   \frac{g}{2\pi}m_{T}\tau_F
   \int_{-\eta_{\max}}^{+\eta_{\max}} d\eta\,\cosh(y-\eta)
 \nonumber\\
   &\times&
   \int_0^{R(\eta)} r_\perp\,dr_\perp \, 
   \I_0\left(\frac{p_\T\sinh\rho(r_\perp)}{T(x)}\right) \\
   &\times&
   \exp\left(\frac{\mu(x)-m_\T\cosh(y{-}\eta)\cosh\rho(r_\perp)}
                  {T(x)}\right) .
 \nonumber 
\end{eqnarray}
where the system is assumed to undergo an instantaneous common freeze-out at a proper time $\tau\equiv\sqrt{t^2-z^2}=\tau_F$. In the above equation, $\eta\equiv\tanh^{-1}(z/t)$ denotes the space-time rapidity and is related to the longitudinal fluid velocity via $\beta_{L}=\tanh(\eta)$. In the transverse plane the flow rapidity (or transverse rapidity) $\rho$ is related to the collective transverse fluid velocity, $\beta_{T}$, via the relation $\beta_{T}=\tanh(\rho)$. 

Considering a Hubble like expansion of the fireball in the transverse plane, a radial dependence of the transverse fluid velocity has been assumed to be of the form
\begin{equation}
\label{beta}
\beta_{T}(r_{\perp}) = \beta^{0}_{T}\left( {\frac{r_{\perp}}{R}} \right)^{n}. 
\end{equation}
where $\beta^{0}_{T}$ is the transverse fluid velocity at the surface of the fireball. The average transverse flow velocity can be easily obtained and is given by $\langle\beta_T\rangle=\frac{2}{2+n}\beta^{0}_{T}$. The transverse flow vanishes at the center and assumes maximum value at the edges, with the flow profile decided by the value of $n$. Most hydrodynamic calculations suggest $n=1$ leading to a Hubble-like transverse rapidity flow profile which is linear in the radial coordinate~\cite{Schnedermann}. Such linear parametrization essentially leads to an exponential expansion of the fireball in the transverse direction, hence the name blast wave.

To account for the limited available beam energy, the longitudinal boost-invariant scenario is modified by restricting the boost angle $\eta$ to the interval  $\eta_{min}\le\eta\le\eta_{max}$. Reflection symmetry about the center of mass ensures $\eta_{min}=-\eta_{max}$ and thus constrains the freeze out volume up to a maximum space-time rapidity $\eta_{max}$. In the transverse plane, the boundary of the fireball is given by $R(\eta)$. Two different choices of $R(\eta)$ are prescribed in Ref.~\cite{Dobler}, corresponding to different shapes of the fireball:
\begin{eqnarray}
   R(\eta) &=& R_0\cdot\Theta\left(\eta_{\max}^2 -\eta^2\right) \,,
 \label{cylinder}\\
   R(\eta) &=& R_0\cdot
   \sqrt{1 - {\eta^2\over\eta_{\max}^2}}\,.
 \label{ellipsoid}
 \end{eqnarray}
The first choice, Eq.~(\ref{cylinder}), describes a cylindrical fireball
in the $\eta-r_{\perp}$-space and corresponds to the usual formalism~\cite{Schnedermann} which was found to work well at top SPS energies and above. However at lower AGS beam energy, the cylindrical symmetry is not fully realized and the fireball is expected to have an elliptic shape~\cite{Nix}, as given by Eq.~(\ref{ellipsoid}). Dependence of transverse size on the longitudinal coordinate explicitly breaks the assumption of boost-invariance.

\begin{figure}
\includegraphics[scale=0.4]{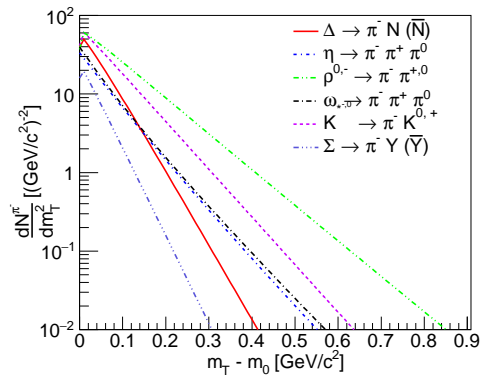}
\caption{An illustration of the resonance decay contributions to the transverse mass spectra of pions. Both two and three body decays are incorporated in the calculation. Higher mass resonances beyond $\Delta(1232)$ are neglected.}
\label{fig1}
\end{figure}

While analyzing the AGS data, the authors of Ref.~\cite{Dobler} had investigated a wide range of possibilities for the different freeze-out parameters. Comparison between fit quality and the number of model parameters (and the related expense in computing time) showed that an optimum description of both longitudinal and transverse spectra can be obtained by reducing the transverse size of the fireball in backward and forward rapidity regions, following Eq.~(\ref{ellipsoid}), along with a temperature and transverse flow gradient constant over the freeze-out surface. In our present analysis, we would, therefore, consider the same parametrization for describing the longitudinal and transverse spectra. 

For comparison to experimental data, one needs to account for hadronic resonance decays. In our present calculations, we follow the formalism given in Ref.~\cite{reso-decay} using thermal distributions Eq.~(\ref{therm}) for the resonances. Both two and three body decay of the sources are numerically simulated. The procedure implies the assumption of full chemical equilibrium which is sufficient for estimating the resonance feed down contributions. The resonance decay contributions to the pion spectra are illustrated in Fig.~\ref{fig1}. We only include hadrons with masses up to $\Delta(1232)$ resonance. As our analysis is restricted up to SPS energies, exclusion of higher resonances would have a negligible effect.


\section{Results and discussions}

\begin{figure*}
\begin{center}
\begin{picture}(200,100)
\put(0,0){\includegraphics[scale=0.3]{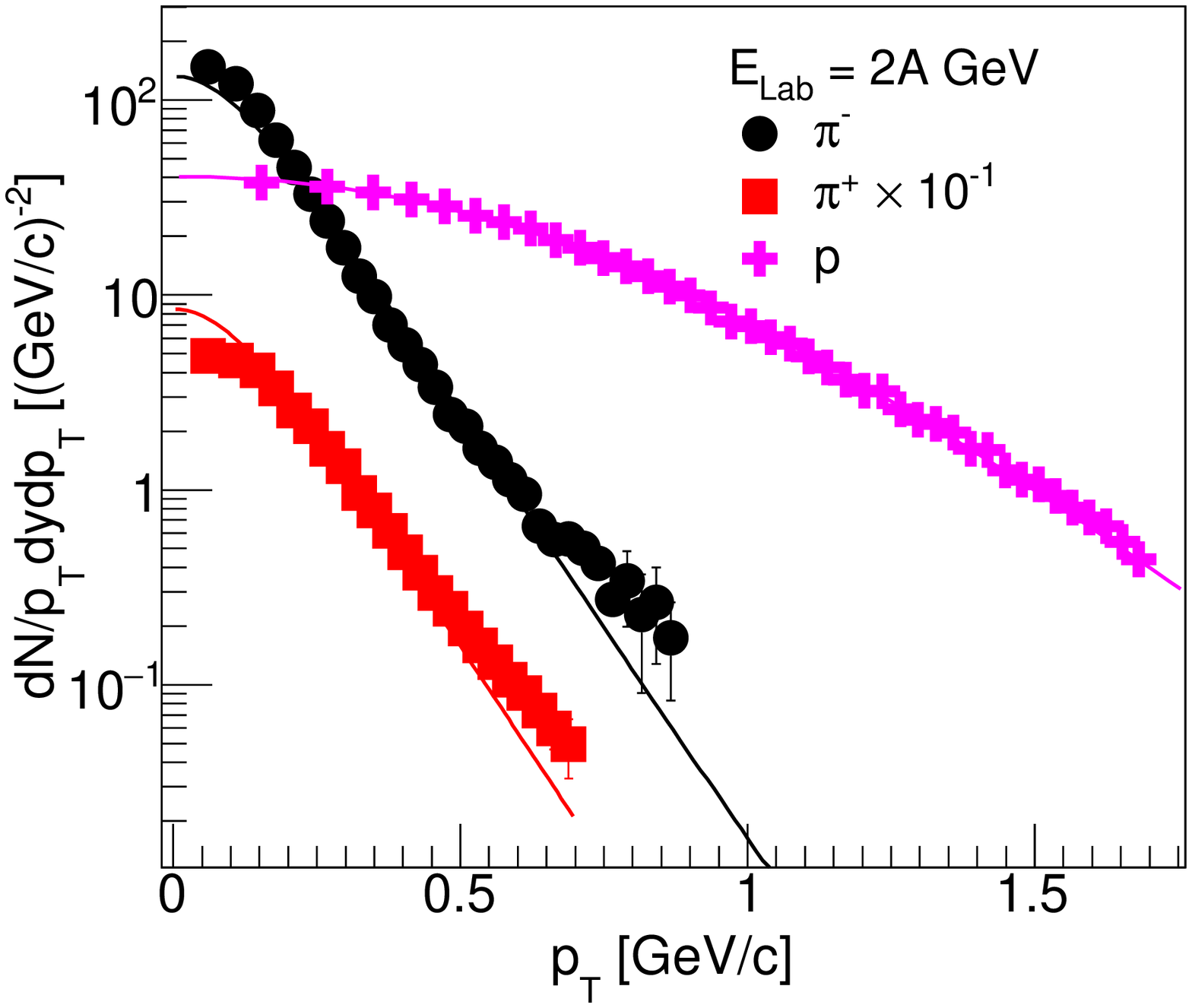}}
\put(50,115){(a)}
\end{picture}
\begin{picture}(200,100)
\put(0,0){\includegraphics[scale=0.3]{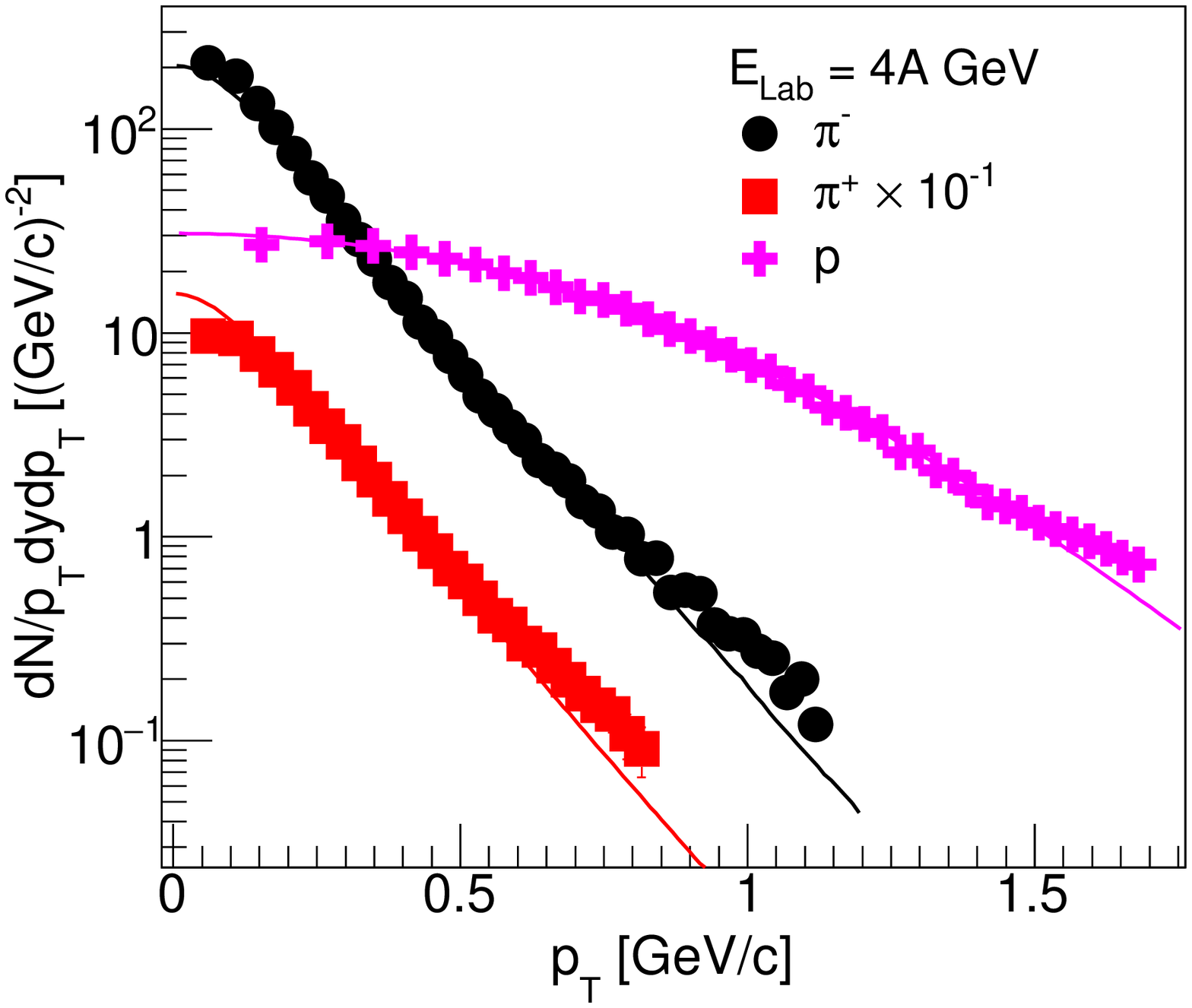}}
\put(50,115){(b)}
\end{picture}\\
\begin{picture}(200,150)
\put(0,0){\includegraphics[scale=0.3]{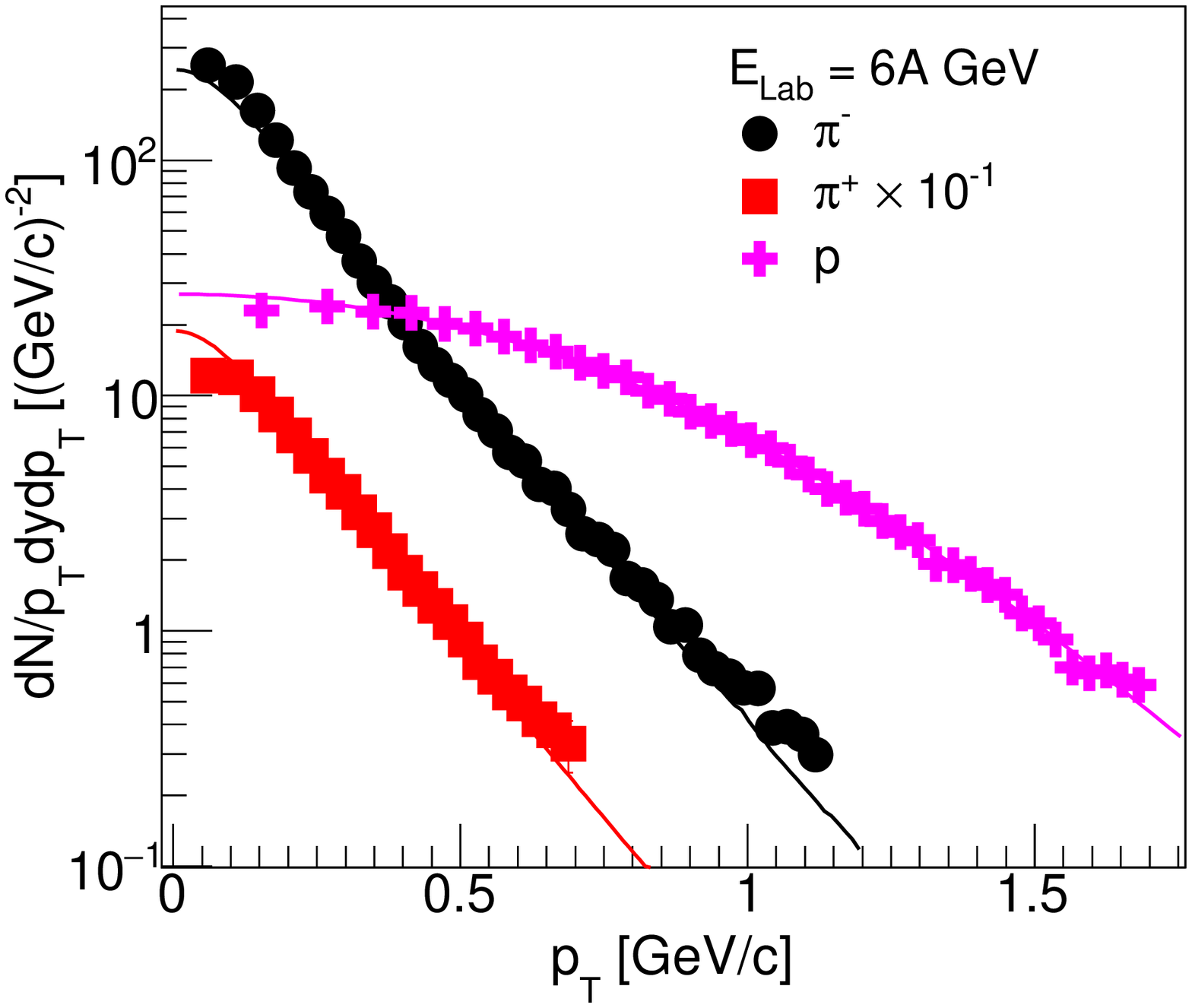}}
\put(50,115){(c)}
\end{picture}
\begin{picture}(200,150)
\put(0,0){\includegraphics[scale=0.3]{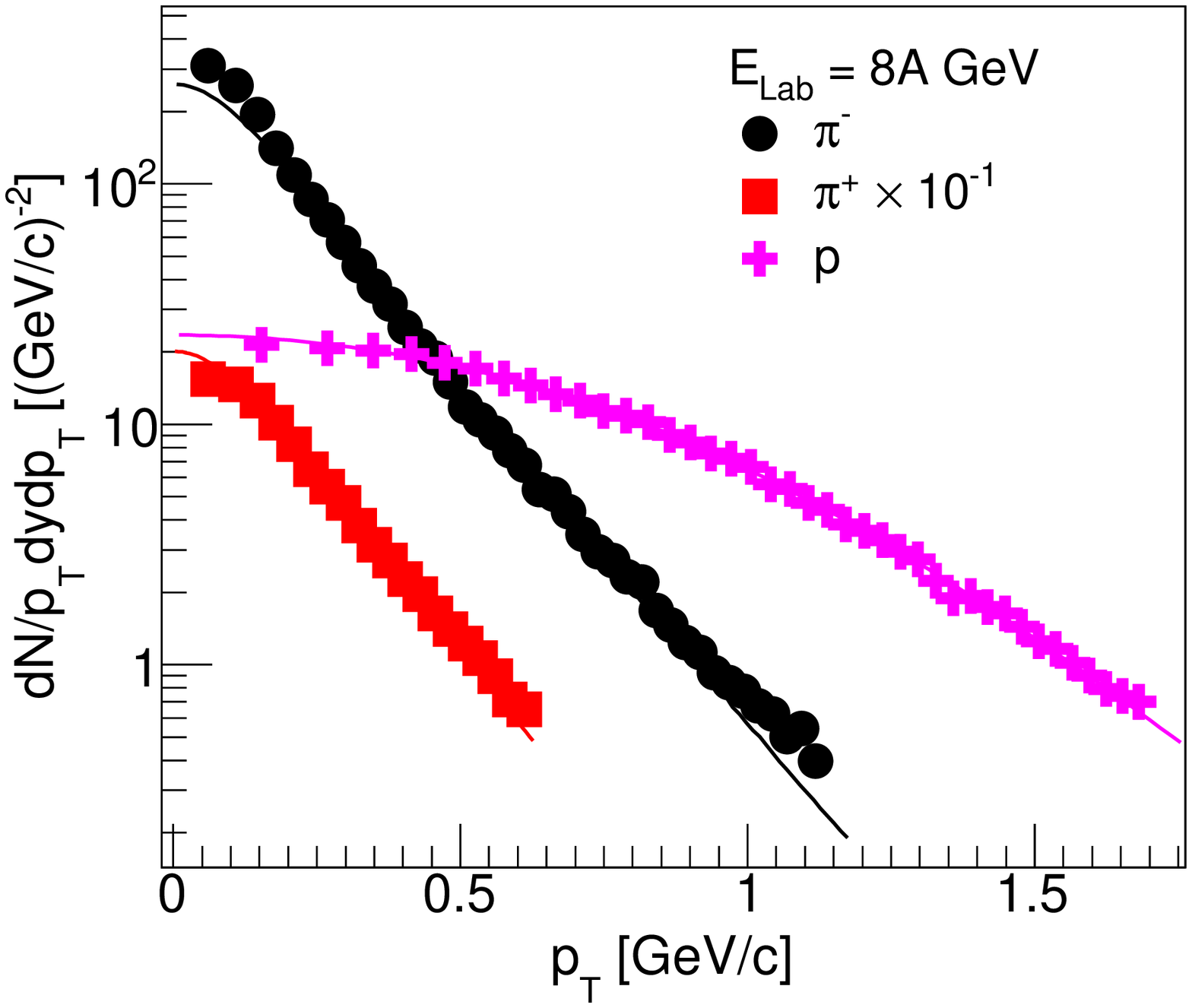}}
\put(50,115){(d)}
\end{picture}
\end{center}
\caption{Fitted $p_{T}$ spectra for Pions ($\pi^{\pm}$) (-0.05 $<$ $y_{c.m.}$ $<$ 0.05) and Proton (p) (-0.05 $<$ $y_{c.m.}$ $<$ 0.05) at (a) 2A GeV, (b) 4A GeV, (c) 6A GeV and (d) 8A GeV beam energies.}
\label{fig2}
\end{figure*}

\begin{figure*}
\begin{picture}(160,140)
\put(0,0){\includegraphics[scale=0.28]{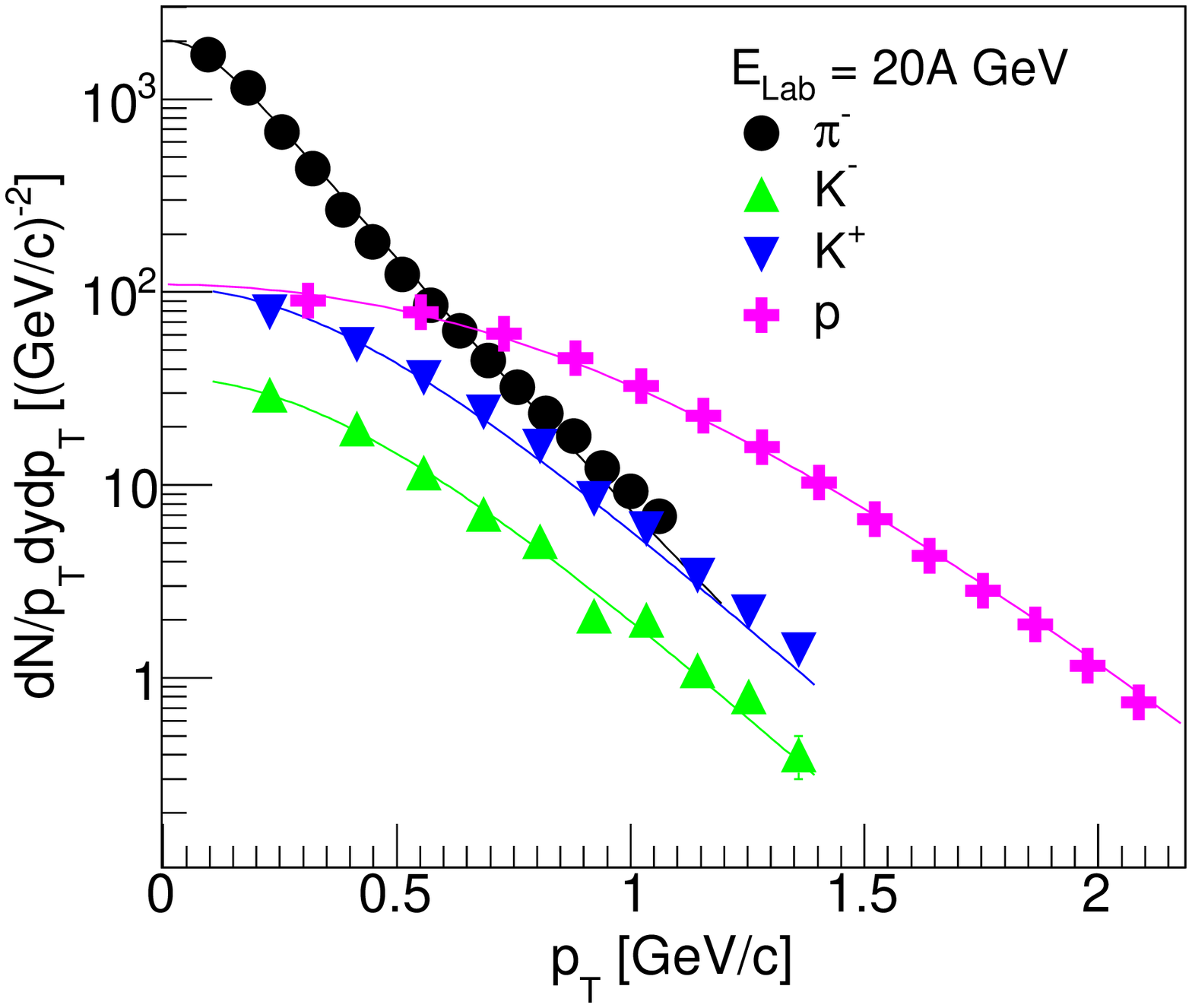}}
\put(50,110){(a)}
\end{picture}
\begin{picture}(160,140)
\put(0,0){\includegraphics[scale=0.28]{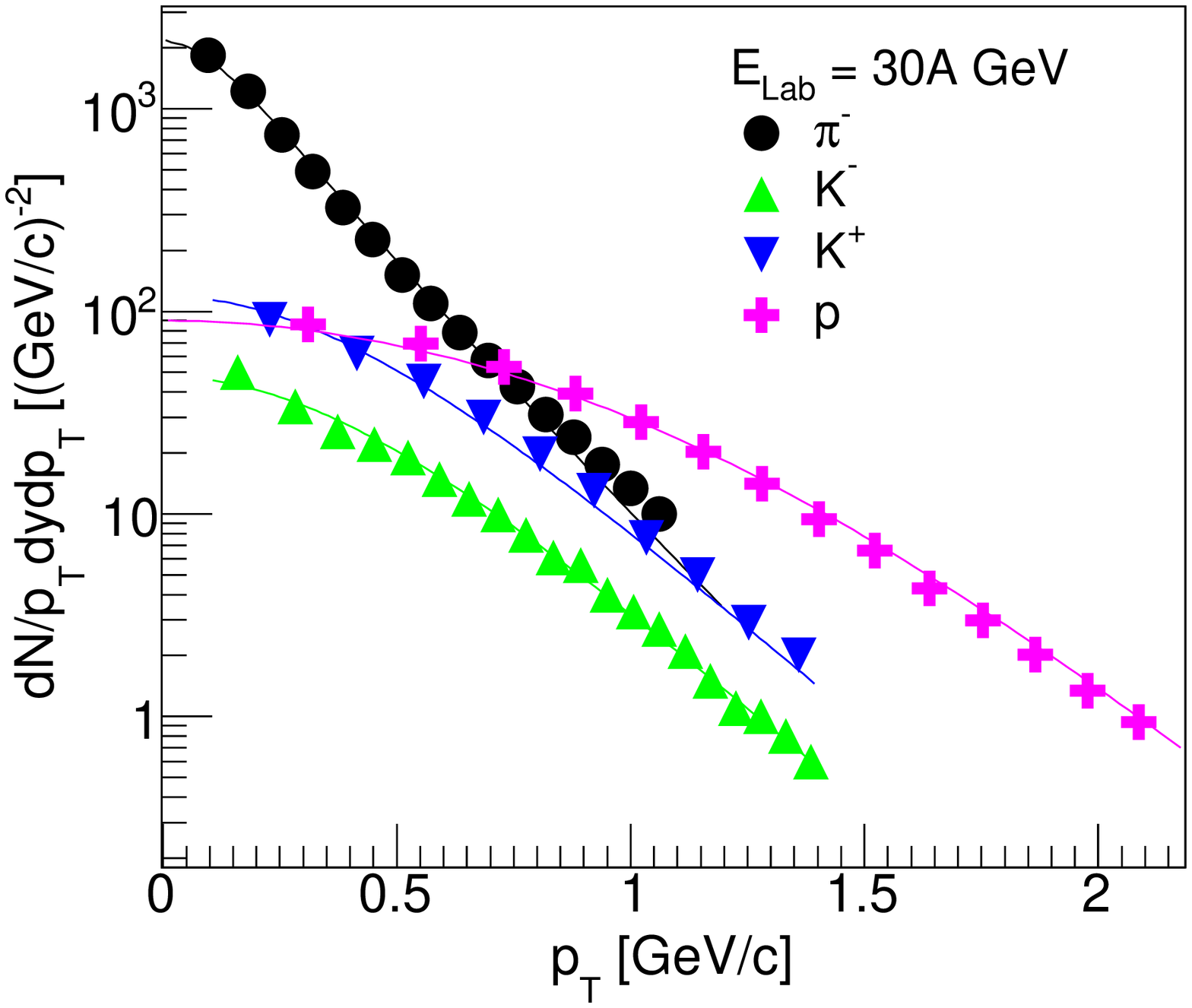}}
\put(50,110){(b)}
\end{picture}
\begin{picture}(160,140)
\put(0,0){\includegraphics[scale=0.28]{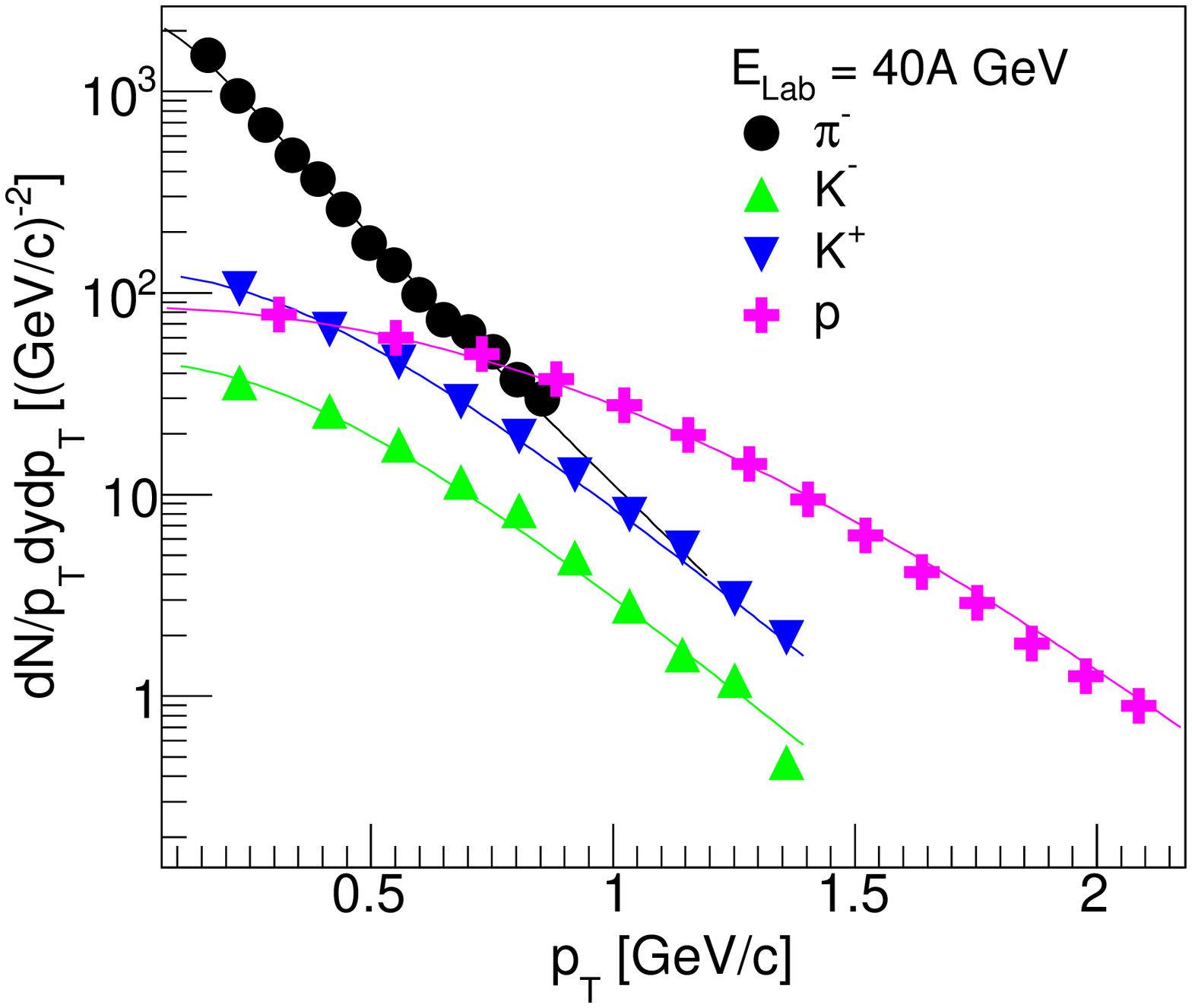}}
\put(50,110){(c)}
\end{picture}
\begin{picture}(160,160)
\put(0,0){\includegraphics[scale=0.28]{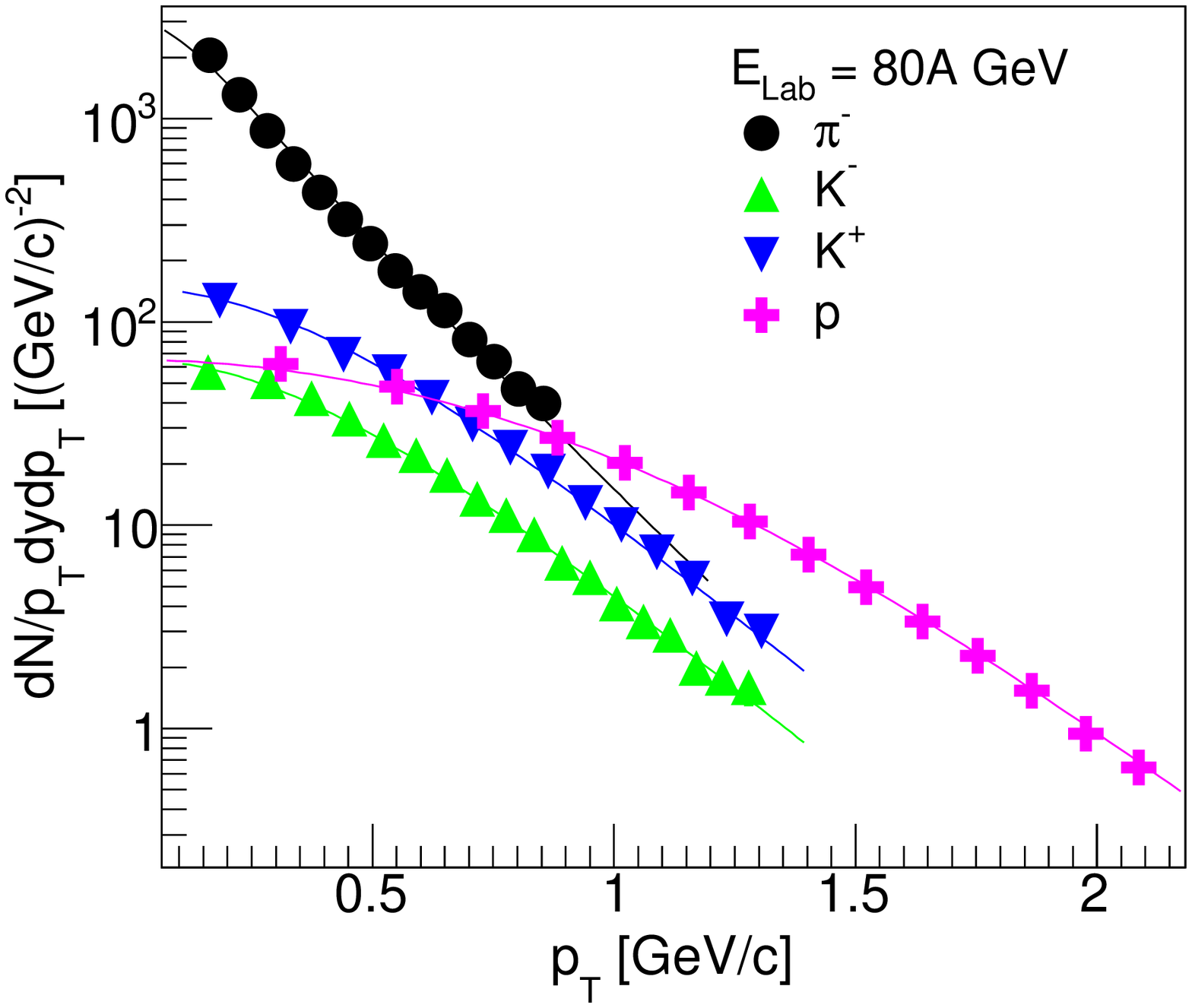}}
\put(50,110){(d)}
\end{picture}
\begin{picture}(160,160)
\put(0,0){\includegraphics[scale=0.28]{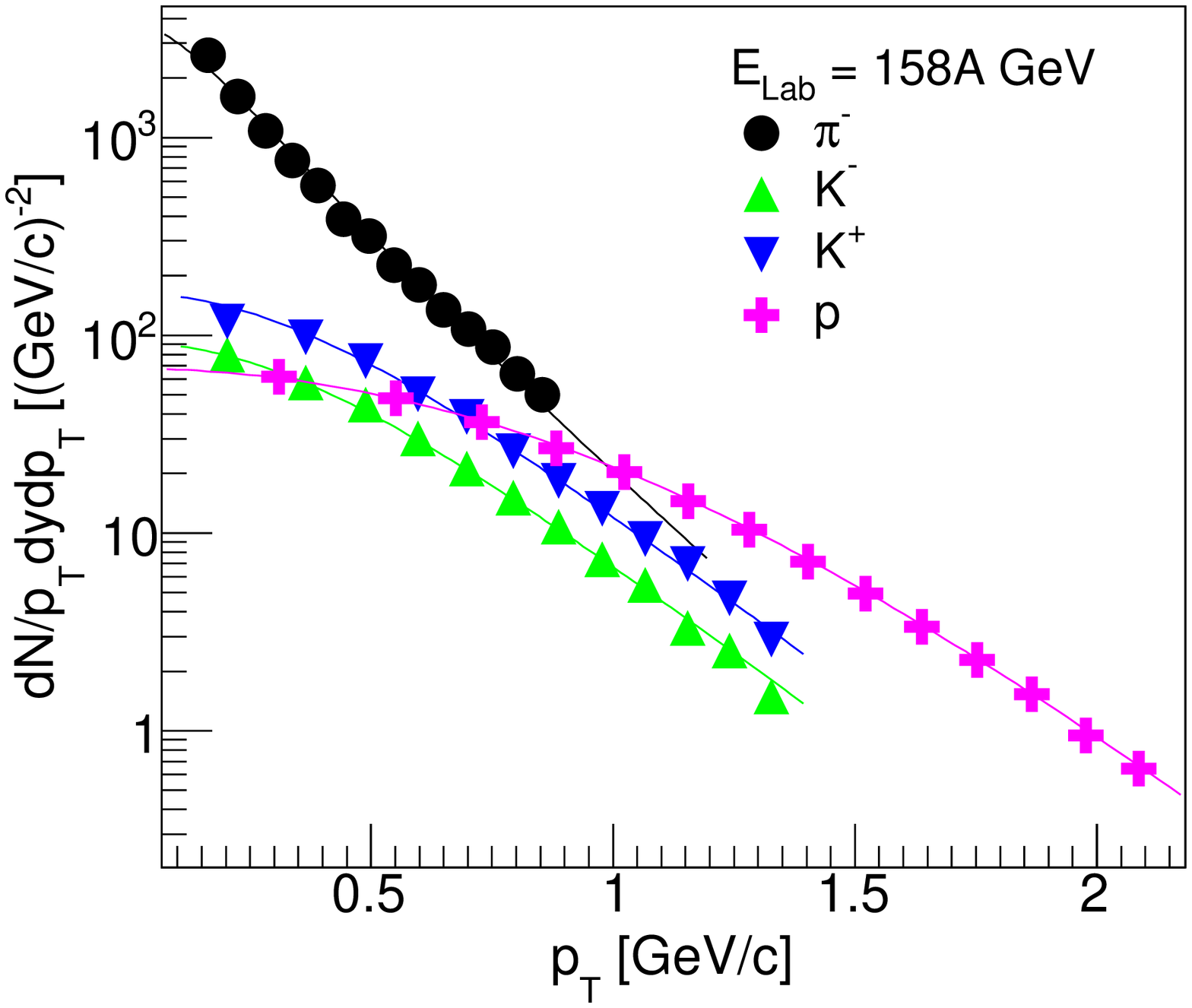}}
\put(50,110){(e)}
\end{picture}
\caption{Fitted $p_{T}$ spectra for  Proton (p)
(-0.38 $<$ $y_{c.m.}$ $<$ 0.32 for 20A GeV, -0.48 $<$ $y_{c.m.}$ $<$ 0.32 for 30A GeV, -0.32 $<$ $y_{c.m.}$ $<$ 0.08 for 40A GeV, -0.36 $<$ $y_{c.m.}$ $<$ 0.04 for 80A GeV and -0.51 $<$ $y_{c.m.}$ $<$ -0.11 for 158A GeV),$\pi^{-}$ (0.0 $<$ $y_{c.m.}$ $<$ 0.2) and $\rm K^{\pm}$ (-0.1 $<$ $y_{c.m.}$ $<$ 0.1) at (a) 20A GeV, (b) 30A GeV, (c) 40A GeV, (d) 80A GeV and  (e) 158A GeV beam energies.}
\label{fig3}
\end{figure*}

\begin{figure*}
\begin{picture}(220,120)
\put(0,0){\includegraphics[scale=0.37]{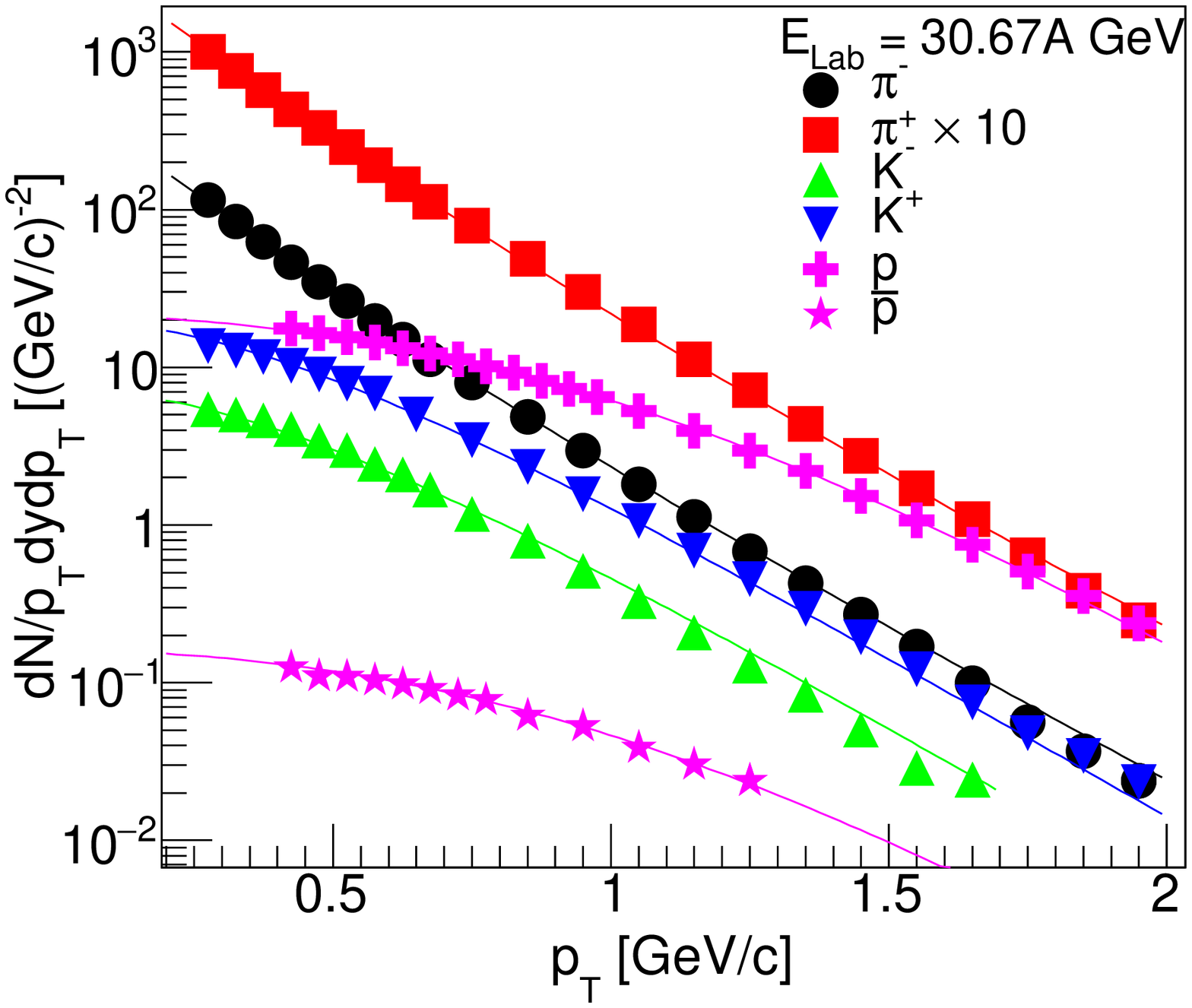}}
\put(90,150){(a)}
\end{picture}
\begin{picture}(220,120)
\put(0,0){\includegraphics[scale=0.37]{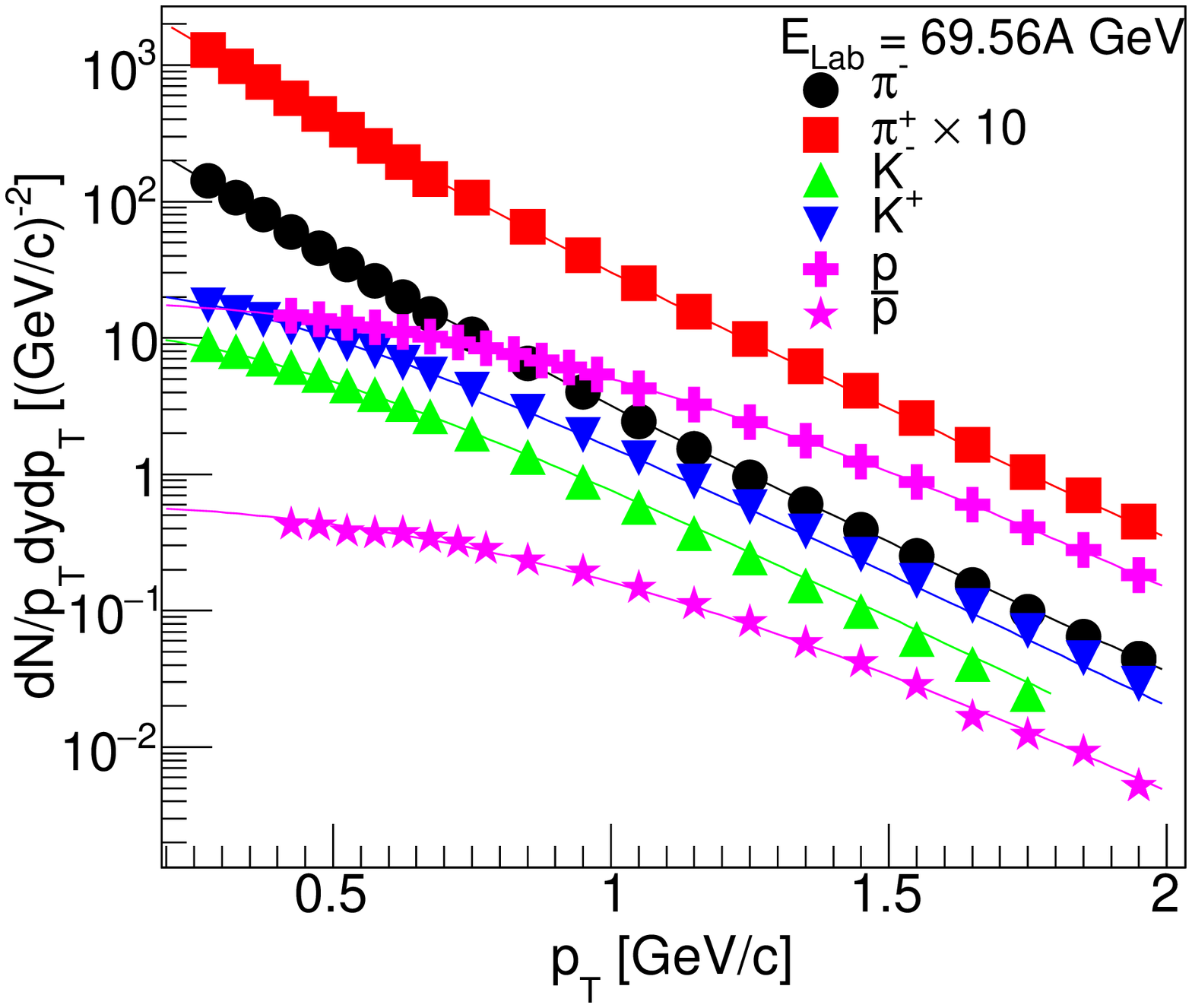}}
\put(90,150){(b)}
\end{picture}
\caption{Fitted $p_{T}$ spectra for  Proton (p) (-0.1 $<$ $y_{c.m.}$ $<$ 0.1), Anti-proton ($\rm \bar{p}$) (-0.1 $<$ $y_{c.m.}$ $<$ 0.1), $\pi^{\pm}$ (-0.1 $<$ $y_{c.m.}$ $<$ 0.1) and $\rm K^{\pm}$ (-0.1 $<$ $y_{c.m.}$ $<$ 0.1) from RHIC beam energy scan (BES) program, at (a) 30.67A GeV and (b) 69.56A GeV beam energies. Since the data have a lower $p_T$ cut off around 0.2 GeV/c, the resonance decay contribution is not included in the calculations.}
\label{fig4}
\end{figure*}

In this section, we present the results of our analysis. For this purpose we consider the measured spectra of the identified hadrons in central Au+Au collisions from E895 Collaboration~\cite{Klay:2001tf, Klay:2003zf} at AGS in the beam energy range $\rm E_{Lab}=2-8 $A GeV and from STAR Collaboration at RHIC BES program~\cite{Adamczyk:2017iwn} for two centre of mass energies $\sqrt{s_{NN}}=7.7$ and $11.5$ GeV ($\rm E_{Lab}=30.67$A and $69.56 $A GeV). In addition data for central Pb+Pb collisions from NA49 Collaboration~\cite{NA4920, NA4940, NA49158} at SPS, in the beam energy range $\rm E_{Lab}=20-158 $A GeV are analysed. We do not go beyond top SPS energy. Note that at AGS the distribution of secondary hadrons was measured by a series of experiments at varying energies and for various collision systems. For the present analysis, we only opt for the latest available data corpus for central Au+Au collisions, from E895 Collaboration. The data were published as acceptance corrected, invariant yield per event as a function of $m_{T}-m_{0}$ ($m_0$ is the particle mass), in small bins of rapidity ($\Delta y=0.1$). For uniformity, in our analysis, we consider only the mid-rapidity bin, where the yield is maximum. For most forward/backward rapidity bins, data points are mostly not available at higher $p_T$. The details of the data sets under investigation, including their beam energy, beam rapidity, collision centrality, phase space coverage and analyzed hadronic species are summarized in Table~\ref{tabI}. As we are interested in the global properties of the fireball, we consider only bulk hadronic species, i.e., $\pi^{\pm}$ and $\rm p$ at AGS energies, $\pi^{\pm}$, $\rm K^{\pm}$, $\rm p$ and $\rm \bar{p}$ at energies available at RHIC beam energy scan (BES) program and $\pi^{-}$, $\rm K^{\pm}$ and $\rm p$ at SPS energies. Due to a lower $p_T$ cut off ($p_{T}^{min}\simeq 0.2$ GeV/c), the resonance decay contribution is excluded while fitting the spectra from RHIC BES program.

\begin{figure*}
\begin{center}
\begin{picture}(200,130)
\put(0,0){\includegraphics[scale=0.34]{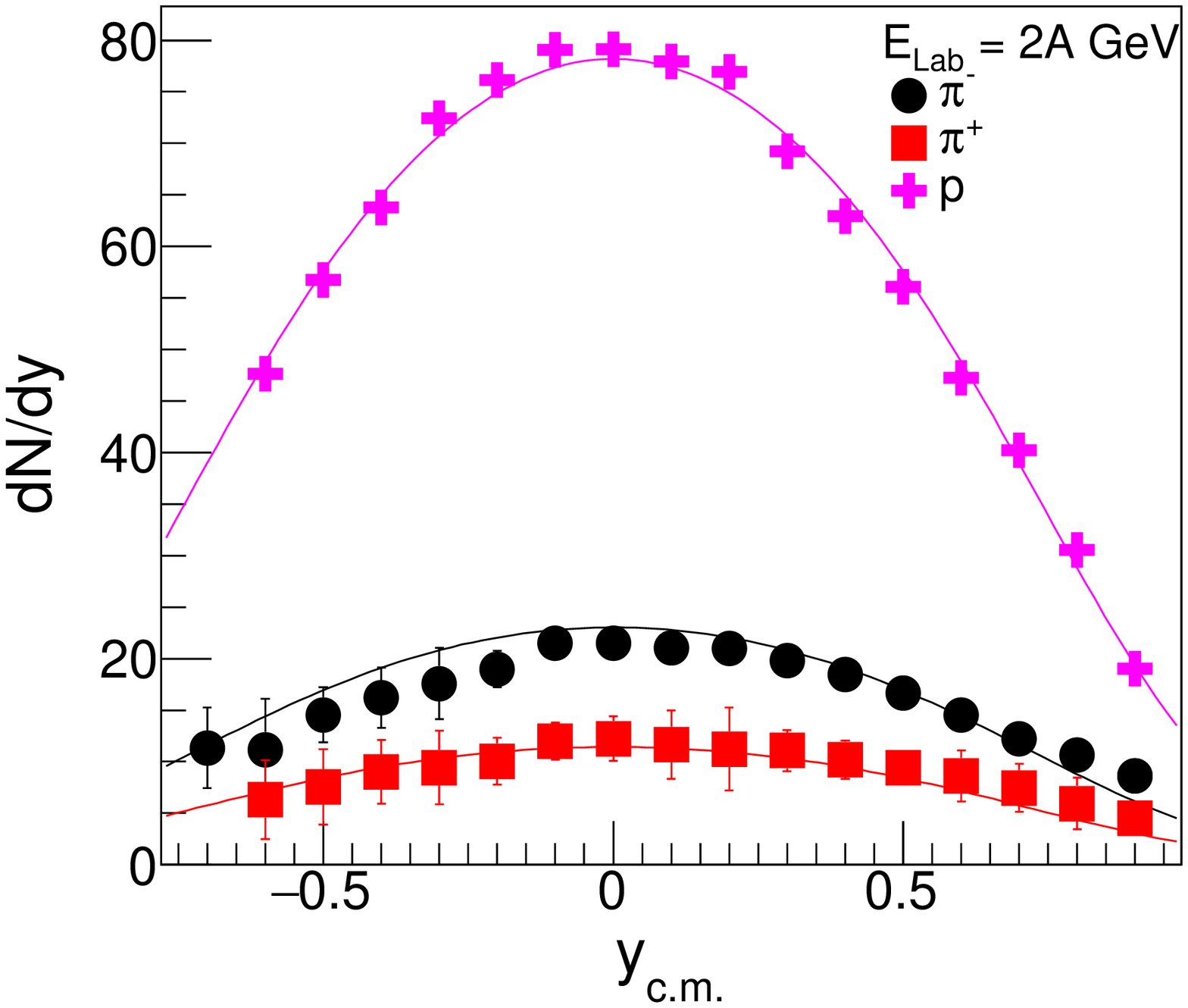}}
\put(35,130){(a)}
\end{picture}
\begin{picture}(200,130)
\put(0,0){\includegraphics[scale=0.34]{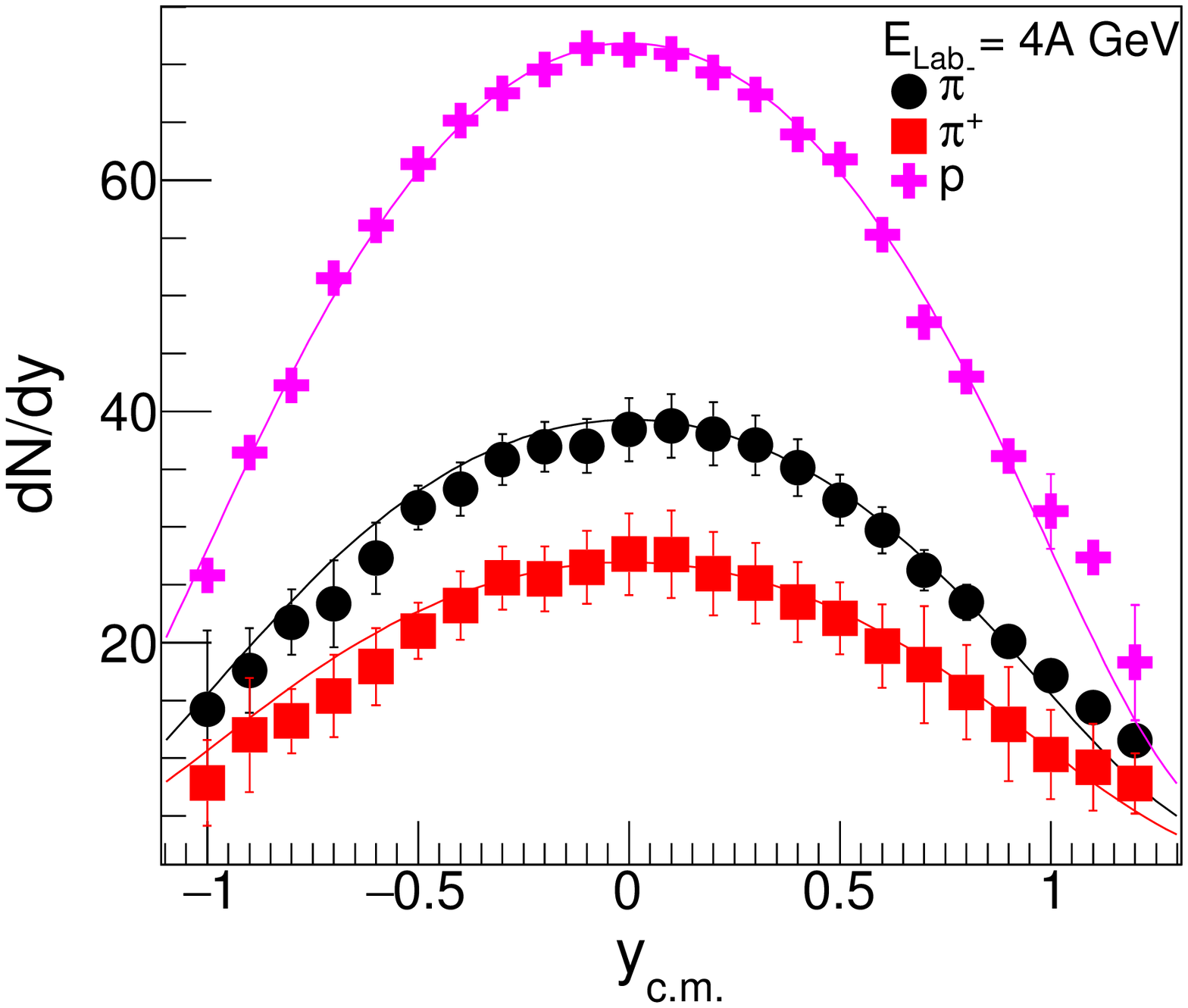}}
\put(35,130){(b)}
\end{picture}
\begin{picture}(200,180)
\put(0,0){\includegraphics[scale=0.34]{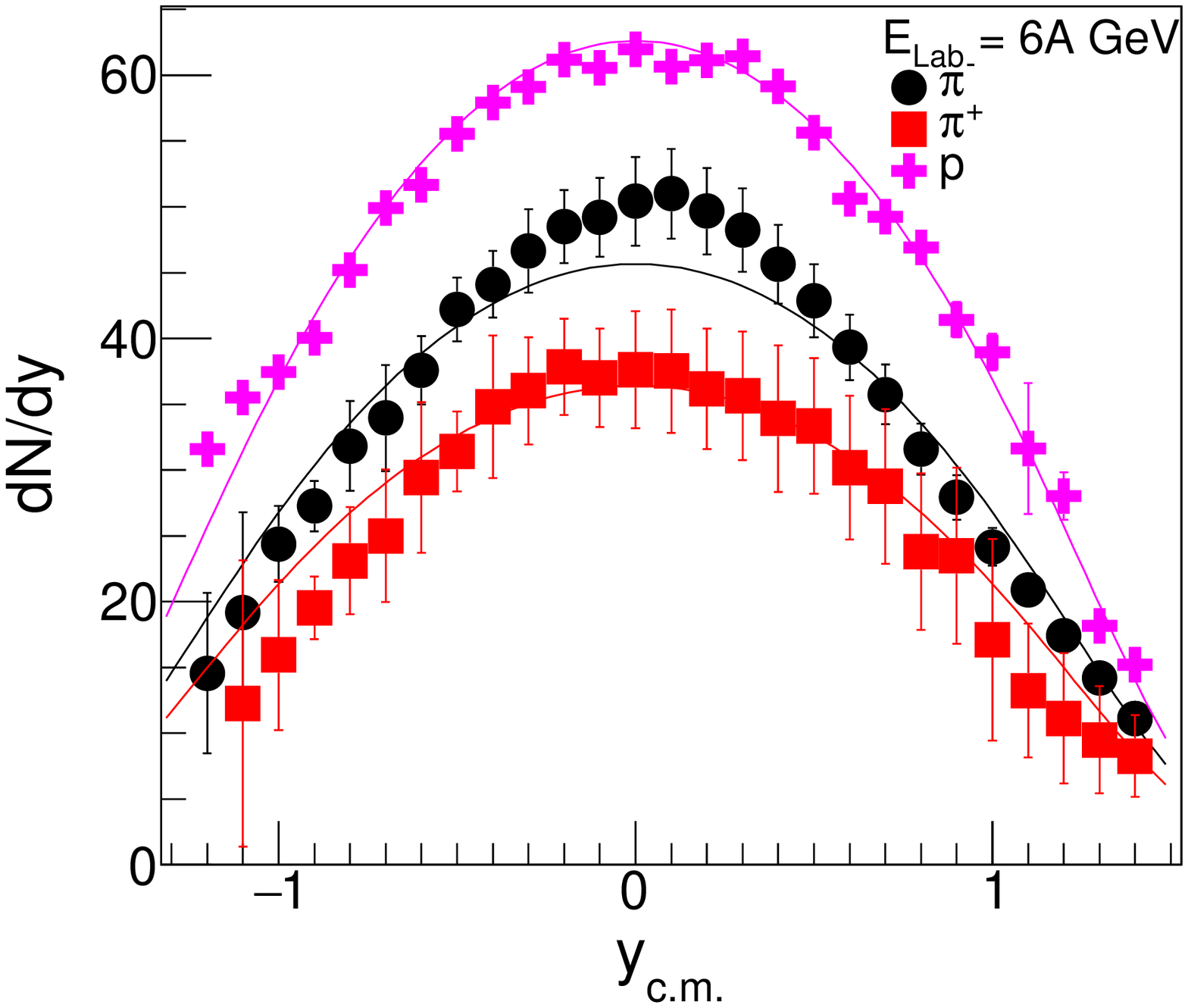}}
\put(35,130){(c)}
\end{picture}
\begin{picture}(200,180)
\put(0,0){\includegraphics[scale=0.34]{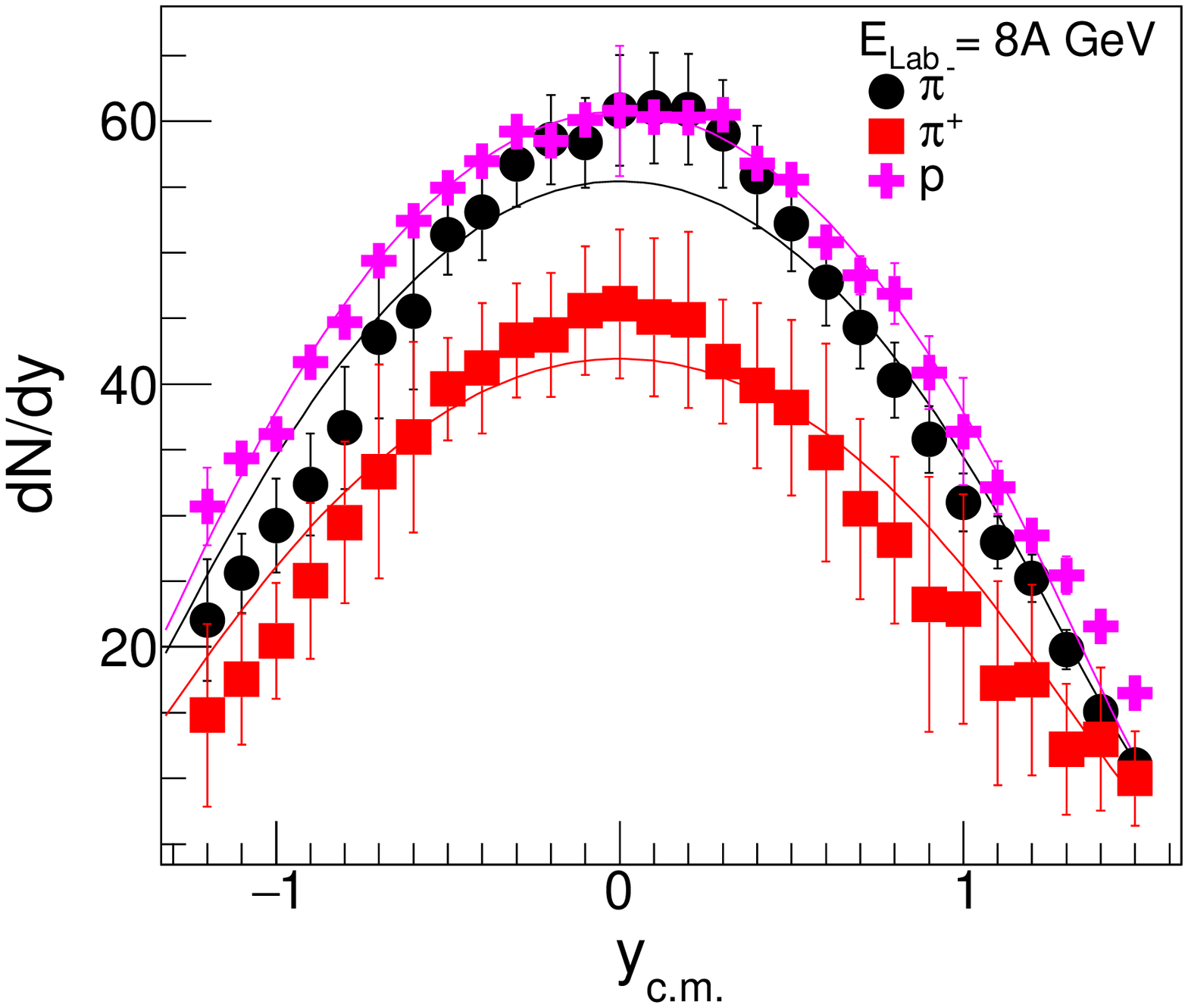}}
\put(35,130){(d)}
\end{picture}
\end{center}
\caption{Fitted rapidity distribution of $\pi^{\pm}$ and Proton (p) in central Au+Au collisions from AGS, at (a) 2A GeV, (b) 4A GeV, (c) 6A GeV and (d) 8A GeV beam energies.
 For each particle species, the normalization constant has been adjusted separately for best fit results.}
\label{fig5}
\end{figure*}

\begin{figure*}
\begin{picture}(160,100)
\put(0,0){\includegraphics[scale=0.28]{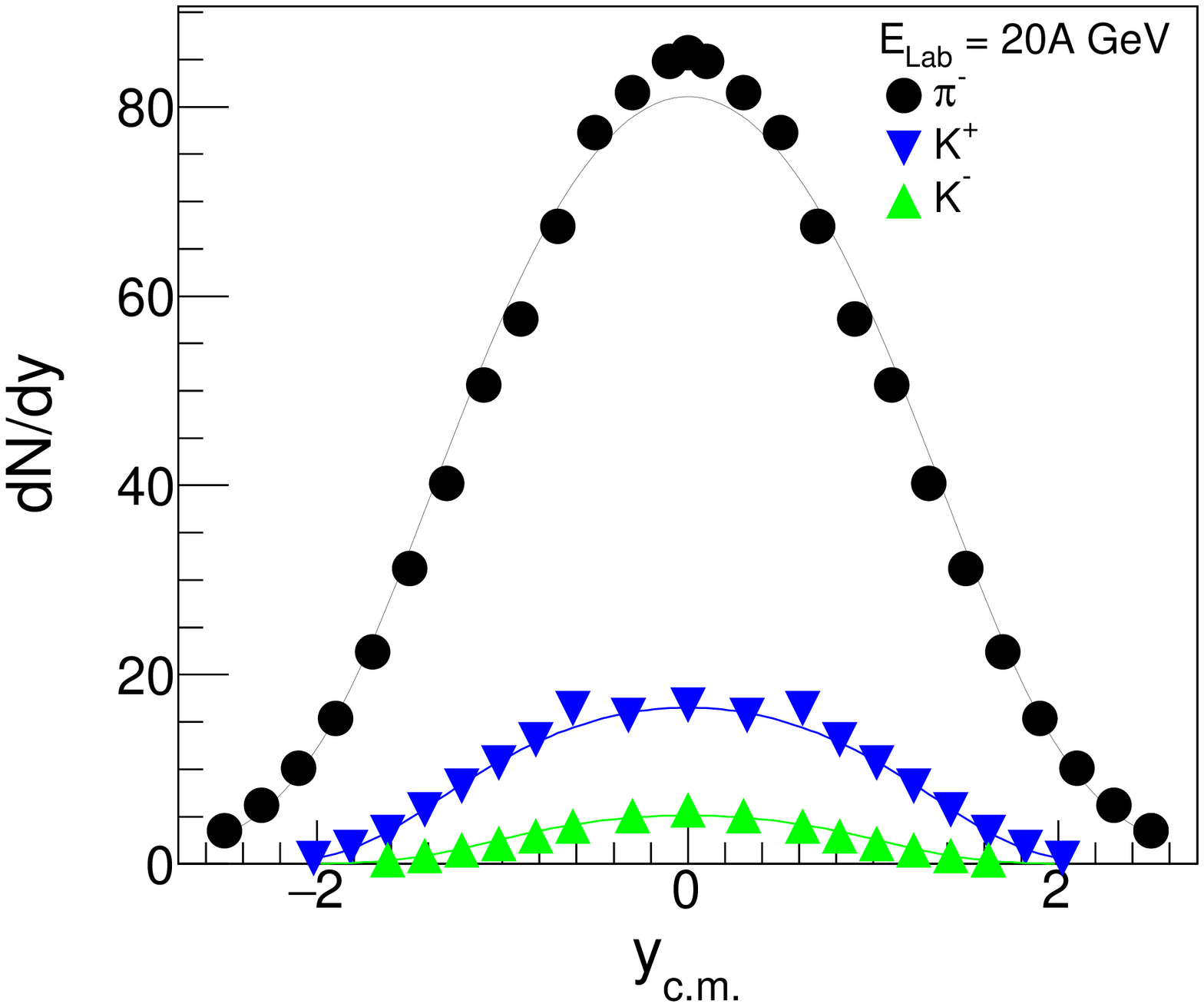}}
\put(30,110){(a)}
\end{picture}
\begin{picture}(160,100)
\put(0,0){\includegraphics[scale=0.28]{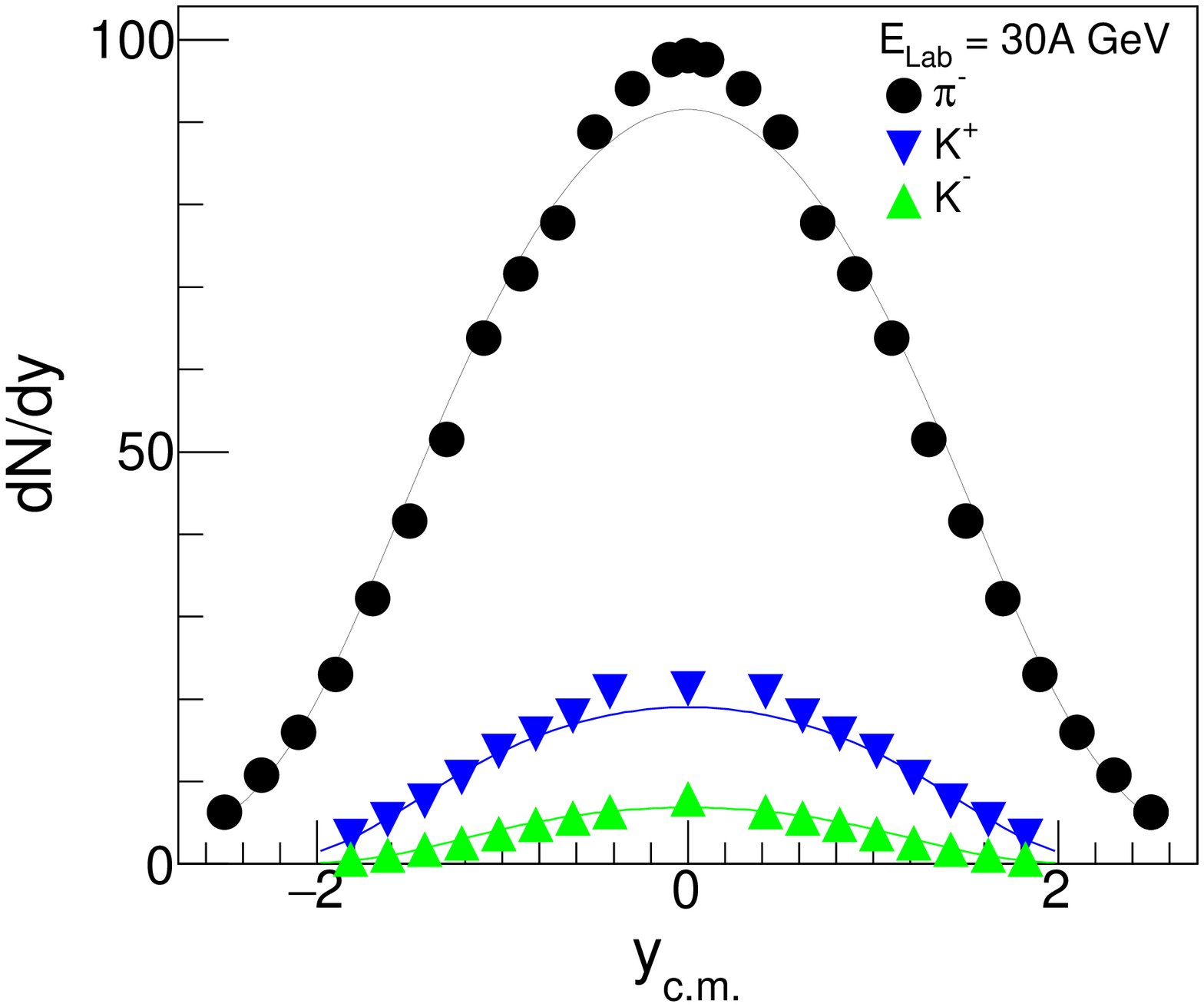}}
\put(30,110){(b)}
\end{picture}
\begin{picture}(160,100)
\put(0,0){\includegraphics[scale=0.28]{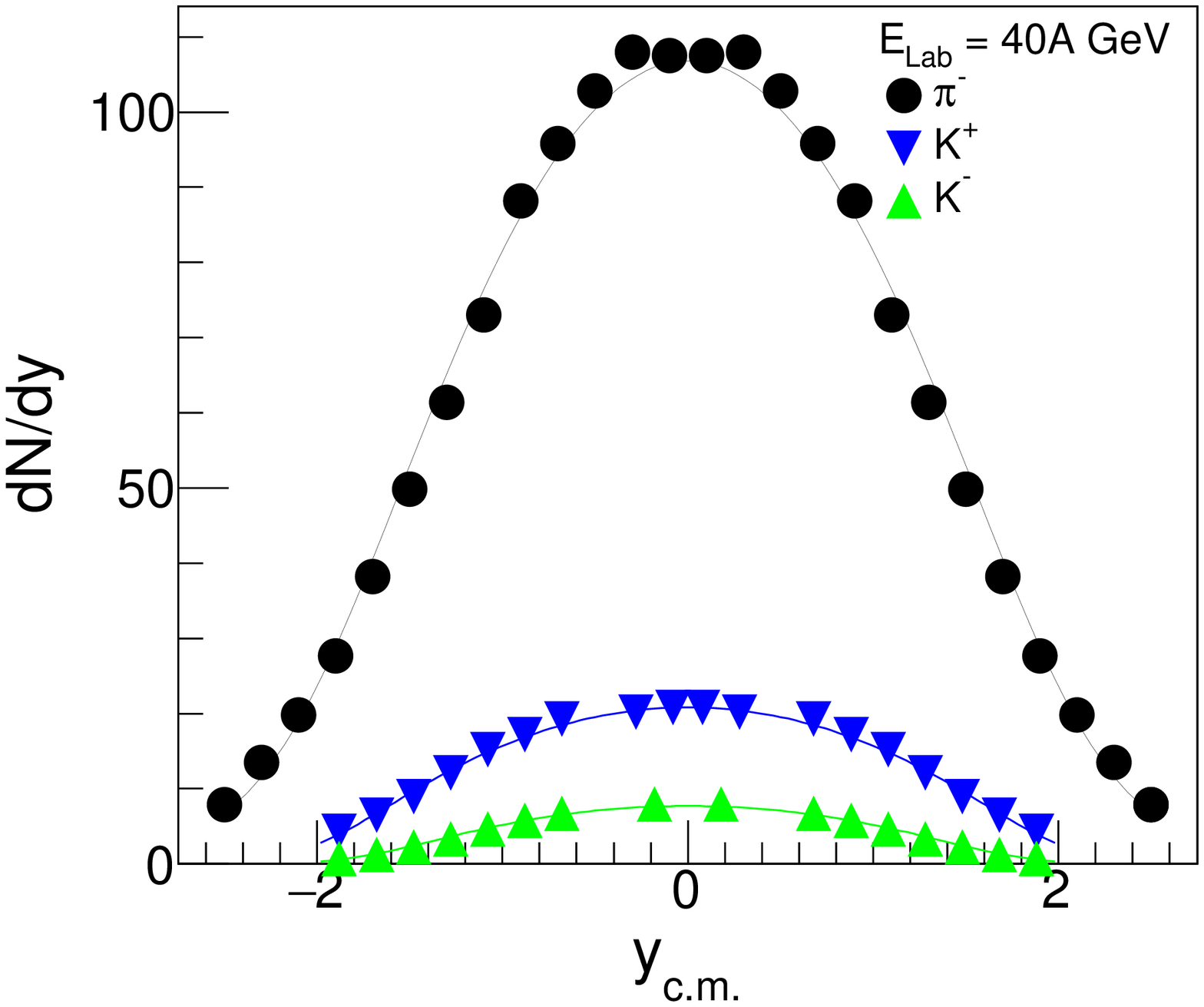}}
\put(30,110){(c)}
\end{picture}
\begin{picture}(160,140)
\put(0,0){\includegraphics[scale=0.28]{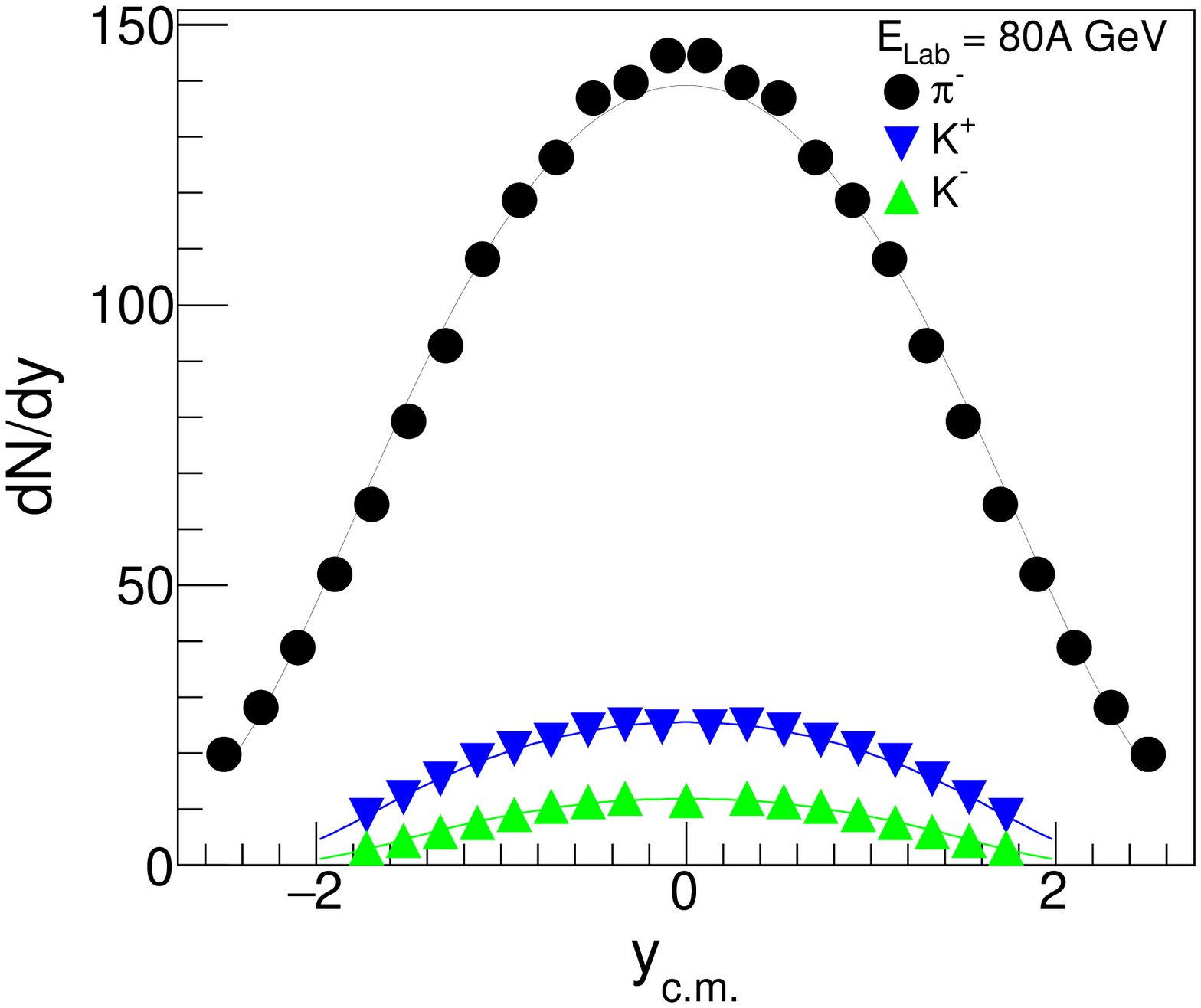}}
\put(30,110){(d)}
\end{picture}
\begin{picture}(160,140)
\put(0,0){\includegraphics[scale=0.28]{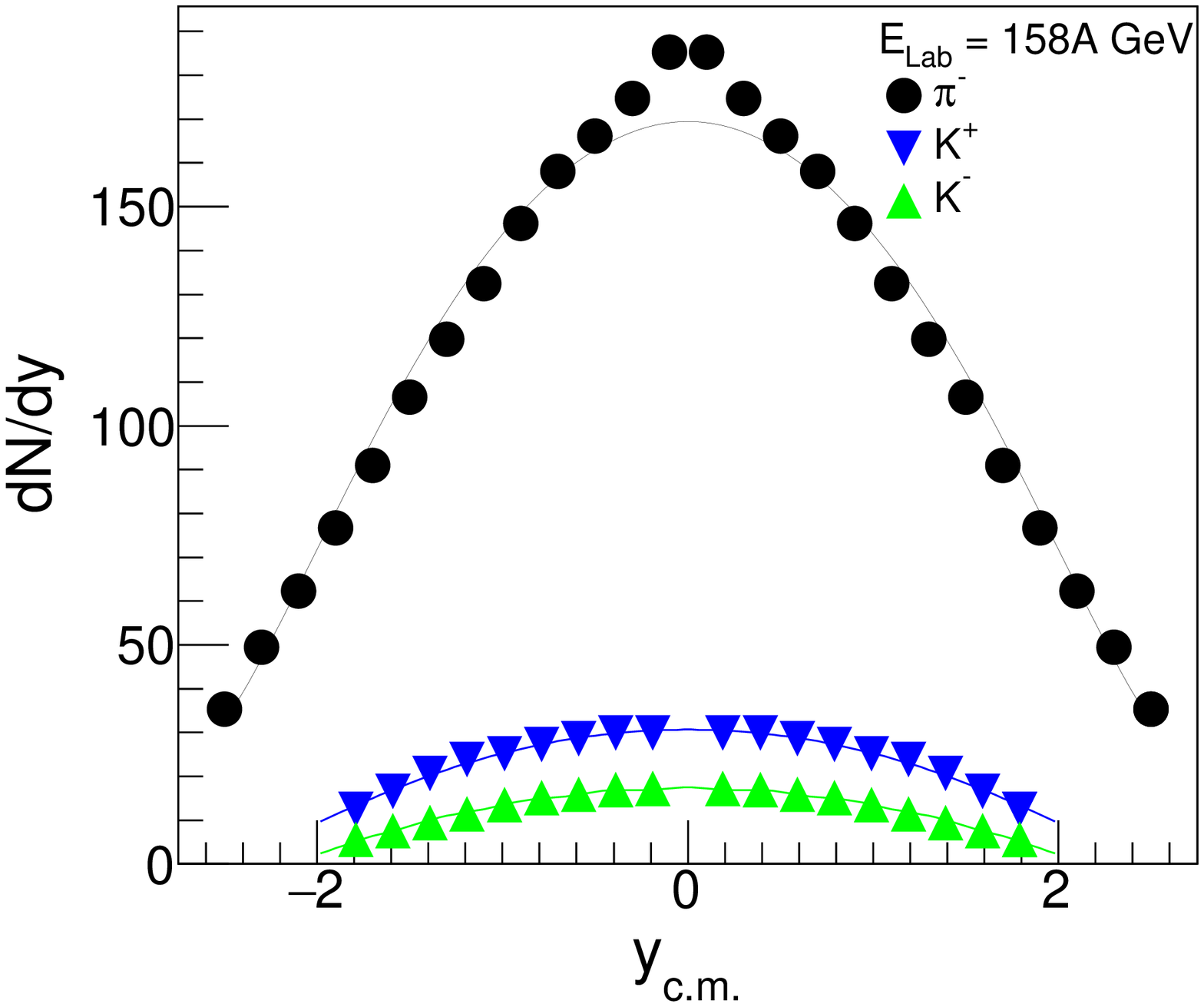}}
\put(30,110){(e)}
\end{picture}
\caption{Fitted rapidity distribution of $\pi^{-}$, $\rm K^{+}$ and $\rm K^{-}$ in central Pb+Pb collisions from SPS, at (a) 20A GeV, (b) 30A GeV, (c) 40A GeV, (d) 80A GeV and  (e) 158A GeV beam energies.}
\label{fig6}
\end{figure*}

We start by fitting the $p_T$ distribution of identified hadrons using Eq.~(\ref{therm}) at different energies. To keep the number of fitting parameters minimal, we couple the freeze-out time $\tau_F$, degeneracy factor $g$ and the fugacity (chemical potential) together into a single normalization constant $Z\equiv{\frac{g}{2\pi}}\tau_F\exp(\mu/T)$, which is adjusted separately for different particle species. Note that the value of chemical potential is fixed at chemical freeze-out and hence its absorption inside the normalization would not affect the thermodynamic conditions at kinetic freeze-out. For a given transverse flow profile ($n=1$), we are thus essentially left with three parameters namely $T$, $\eta_{max}$ and $\beta^{0}_{T}$ which are common for all hadrons at a given energy and extracted from the simultaneous fitting of the $p_T$ spectra of selected hadronic species. The extracted parameters are then used to describe the rapidity spectra of those particles. The resulting $p_T$ spectra at AGS, SPS and RHIC BES are shown in Fig.~\ref{fig2}, \ref{fig3} and \ref{fig4} respectively. The corresponding best fitted values of the parameters $\eta_{max}$, $\langle\beta_{T}\rangle$ and $T_{kin}$ at different beam energies, obtained via minimization of reduced $\chi^{2}$ (defined as $\chi^2$ per degree of freedom) are given in the Table~\ref{tabII}. 
 
As evident, the model gives a reasonable description of the $p_{T}$ spectra of the bulk hadrons at all investigated energies. The freeze-out temperature is found to be relatively low which gradually increases with beam energy. Rather a strong transverse collective motion is observed even at lowest AGS energy. Hadronic $p_T$ spectra, in this investigated energy domain, has also been analyzed within the boost-invariant blast wave model. Relatively higher freeze-out temperatures ($T_{kin}>100$ MeV) has been observed even at AGS energies with a slightly weaker transverse flow~\cite{Sandip}. However one should take note of the fact, that in the corresponding analysis particles are chosen above a non zero $p_T$ value (eg: 0.5 GeV/$c$ for pions) to exclude the effect of resonance decay. Also, the transverse flow parameter $n$ is kept free (which is about 0.5) and fixed from the data whereas we set it to $n=1$. 

In this context, we would also like to mention that at AGS, previous attempts have also been made to fit the transverse distribution of the hadrons with a static rapidity dependent two slopes empirical model in absence of any collective flow~\cite{Klay:2003zf}. The two inverse slope parameters $T_{1}$ and $T_{2}$ respectively dominate the low and high end of the $m_{T}-m_{0}$ spectra. Both $T_{1}$ and $T_{2}$ assume maximum values at midrapidity, with $T_{1}$ around $50$ MeV and $T_{2}$ around $130$ MeV.

\begin{figure}
\includegraphics[scale=0.4]{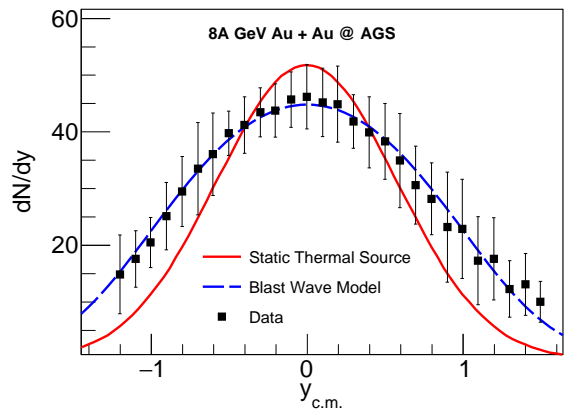}
\caption{Rapidity density distributions of pions in 8A GeV central Au+Au collisions at AGS. Data are compared with predictions from a static thermal model and non boost-invariant blast wave model.}
\label{fig7}
\end{figure}

\begin{figure}
\includegraphics[scale=0.44]{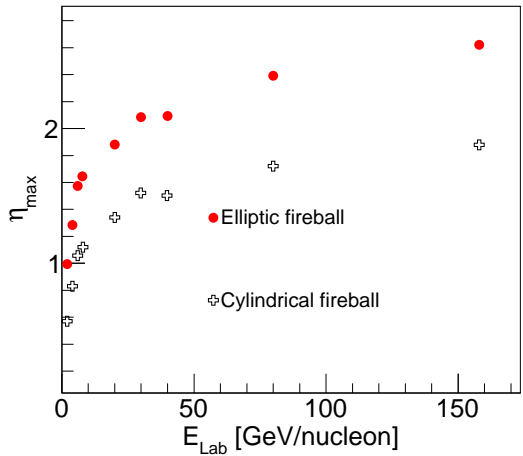}
\caption{Comparison of the $\eta_{max}$ values for different beam energies at AGS and SPS for an elliptic fireball and a cylindrical fireball. The value $\eta_{max}$ is consistently larger for former case compared to the latter one.}
\label{fig8}
\end{figure}

After a successful description of the $p_T$ spectra, we now move on to the description of the rapidity distribution of the produced hadrons. Longitudinal spectra are useful to explore the collective effects in the longitudinal direction. Integrating Eq.~(\ref{therm}) over the transverse components we obtain the rapidity distribution which is contrasted with the available data from AGS and SPS. No data on rapidity distribution of the bulk hadrons is available from RHIC BES program. The same $\eta_{max}$ values as listed in Table~\ref{tabII} are now used to describe the rapidity density distributions. However, the freeze-out temperatures $T_{kin}$ are not chosen from the fit to the $p_T$ spectra. The $p_T$ spectra are fitted only in the mid-rapidity region. In a non boost invariant model, the physical quantities, including the temperature would depend on the rapidity of the measured hadrons. One might also recall that rapidity spectra are rather insensitive to the underlying temperature~\cite{Schnedermann}. Change of temperature has a minimal effect on the corresponding value of $\eta_{max}$. Temperature values as high as $120 - 150$ MeV can also describe the observed rapidity distributions of the produced particles. In our calculations, we fix $T_{kin}$ to 120 MeV which gives a reasonable fit to the rapidity spectra. However such high values of temperature cannot provide a reasonable fit to the $p_T$ spectra. The results are depicted in Fig.~\ref{fig5} and \ref{fig6}.
 
The rapidity distribution of a particle of mass $m$, emitted from a static thermal source at temperature $T$ has the form
\begin{align} 
\label{dndythermal}
{{dn_{\rm th}}\over{dy}} =&~ { V \over {(2\pi)^2} } T^3
    \left(
    {{ m^2 }\over{T^2}}  + {m\over T} {2\over{\cosh y}}
                + {{2}\over{\cosh^2 y}} \right)\nonumber\\
    &\times\exp\left(-{m\over T} \cosh y \right) \; 
\end{align}
where $V$ denotes the source volume. It is well known that the measured particle rapidity distributions from experiments at all beam energies cannot be described by isotropic emission from static thermal models; observed distributions being much wider compared to model predictions. Thermal models incorporating collective expansion in the longitudinal direction, have been much more successful in reproducing the observed rapidity distributions. 

\begin{figure}
\includegraphics[scale=0.44]{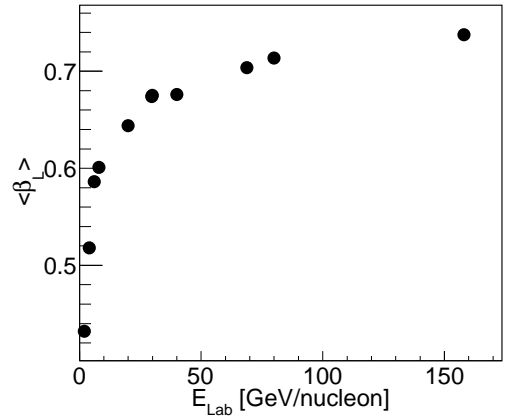}
\caption{Variation of the average longitudinal velocity of the fireball with beam energy.}
\label{fig9}
\end{figure}

An illustrative comparison is presented in Fig.~\ref{fig7}, where rapidity distribution of the pions in 8A GeV central Au+Au collisions is contrasted with that from a static thermal model as well as from the present blast wave model calculations. As evident rapidity distribution from a static isotropic thermal source falls much faster than the data. A similar feature is observed for all other particles and at all the investigated energies. The additional collective motion is attributed to the large pressure gradients developed inside the hot and dense nuclear matter fireballs created in the early stage of the collisions. Note that the inclusion of longitudinal expansion is generally carried out with a longitudinally boost-invariant superposition of multiple boosted individual sources, locally thermalized and isotropic, in a given rapidity interval~\cite{Netrakanti}. Each locally thermalized source is modeled by the $m_T$-integrated Maxwell-Boltzmann distribution, with the rapidity dependence of the energy, $E=m_T \cosh y$ explicitly included. Thus within a boost-invariant scenario, the rapidity distribution from a boosted thermal source can be written as
\begin{equation}
\label{dndythint}
{{dn}\over{dy}}(y) =
    \int\limits_{-\eta_{\rm max}}^{\eta_{\rm max}} \! d\eta \,
        {{dn_{\rm th}}\over{dy}} (y-\eta)
\end{equation}
Note that Eq.~(\ref{dndythint}) is equivalent to what we obtain by integrating Eq.~(\ref{therm}) over $p_T$, for a cylindrical fireball as given by Eq.~(\ref{cylinder}). For comparison, we independently fit the rapidity distribution of pions for both $\eta$ dependent and $\eta$ independent transverse radius of the fireball. A comparison of the extracted values of $\eta_{max}$ at different energies, for elliptic fireball and cylindrical fireball, is displayed in Fig.~\ref{fig8}. For the non boost-invariant model $\eta_{max}$ (and hence the maximum longitudinal fluid velocity) is consistently higher than a boost-invariant case. 

In closing it might also be interesting to investigate the effect of longitudinal flow within the present model. For a cylindrical fireball, in the flat portion of the rapidity distribution, one can define an average longitudinal velocity as $\langle\beta_{L}\rangle = \tanh(\eta_{max}/2)$ \cite{Netrakanti}. Note that the longitudinal velocity $\beta_{L}$ is not a linear function of $\eta$ and therefore the above expression for $\langle\beta_{L}\rangle$ may not hold for non boost-invariant models. For non boost-invariant models, the longitudinal and transverse motion of the thermal source can no longer be decoupled. The rapidity distribution is no longer flat in $\eta$ within the region $\eta_{min}<\eta<\eta_{max}$. One can still define an average of the magnitude of longitudinal velocity of the medium as 
\begin{equation}
\begin{centering}
\label{betaL}
\langle\beta_{L}\rangle = {{\int_{0}^{\eta_{\max}} d\eta\,\tanh(\eta)} 
 \over {\int_{0}^{\eta_{\max}} d\eta}} = \frac{\ln(\cosh\eta_{\max})}{\eta_{\max}}.
\end{centering}
\end{equation}
From the above expression, we see that $\langle\beta_{L}\rangle\to 1$ as $\eta_{max}\to\infty$. Note that, as $\eta_{max}\to\infty$, longitudinal boost-invariance symmetry is restored. The dependence of $\langle\beta_{L}\rangle$ with $E_b$ is shown in Fig.~\ref{fig9}. We observe that $\langle\beta_{L}\rangle$ initially increases and gradually shows a saturative trend with increasing beam energy. This is a consequence of the result shown in Fig.~\ref{fig8} where we find that the fitted value of $\eta_{max}$ increases as the beam energy is increased. Therefore, from our analyses of beam energy dependence of $\eta_{max}$ and $\langle\beta_{L}\rangle$, we conclude that longitudinal boost-invariance is expected to be recovered at high collision energies. 

One may also note that since protons are present before the collision, they are subject to nuclear transparency effect, which is not incorporated in our calculations. At higher beam energies this effect might be strong, which can also broaden the proton rapidity distributions. However, we do not use only protons to determine the absolute magnitude of the collective motion. Rather we compare multiple particle species from the same collisions simultaneously to determine the degree of collectivity, and hence the effect of nuclear transparency is expected to be small.


\section{Summary}

In this paper, we have studied the kinetic freeze-out conditions of bulk hadrons in central Au + Au and Pb + Pb collisions, at AGS, SPS and partially at RHIC BES energies, using a non boost-invariant version of the blast wave model. The assumption of boost-invariance is explicitly broken by introducing a dependence of the transverse size of the fireball on the space-time rapidity. The double differential transverse momentum spectra for a variety of particle species are simultaneously analyzed. The overall fit to the data is reasonably good over a wide range of beam energy. The results indicate a relatively low $T_{kin}$ in the range $55-90$ MeV with a substantial $\langle\beta_{T}\rangle$ of about $0.55c-0.6c$. We also found that $T_{kin}$ increases gradually with the incident beam energy. To explore the effect of longitudinal dynamics, the available rapidity spectra were also analyzed. The $\eta_{max}$ values showed a monotonically increasing trend from AGS to SPS energies. Higher values of $\eta_{max}$ were observed in case of the elliptic fireball than the cylindrical one. This may be attributed to the fact that one needs a larger value of $\eta_{max}$ for ellipsoidal cross-section compared to the cylindrical one, in order to have an identical volume of the fireball needed to reproduce the measured rapidity spectra. For the upcoming experiments at FAIR and NICA accelerator facilities, these measurements would be useful to better understand the freeze out conditions. 

In addition to two freeze-out scenarios, with the chemical freeze-out preceding the thermal freeze-out, the so-called single freeze-out models are also available in the literature, both for boost-invariant~\cite{sfo} as well as non boost-invariant longitudinal dynamics~\cite{Bearden:2004yx}. In these models, the hadron spectra are affected by the transverse expansion as well as the decay of resonances in such a way that it is possible to describe well both the particle spectra and the particle ratios with a single value of the temperature. Accounting for the hadronic decays essentially leads to an effective cooling of the spectra. It will be interesting to analyze the present dataset within the single freeze-out scenario in the future.


\begin{acknowledgements}
We would like to thank Premomoy Ghosh, Sushant Singh and Ralf Averbeck for critically reading the manuscript. P. P. B. is grateful to Kai Schweda, Anton Andronic and Premomoy Ghosh for illuminating discussions on blast wave model. A.J. is supported in part by the DST-INSPIRE faculty award under Grant No. DST/INSPIRE/04/2017/000038.
\end{acknowledgements}



\end{document}